\documentclass{article}

\usepackage{cite}
\usepackage{amsmath,amssymb,amsfonts}
\usepackage{orcidlink}
\usepackage{authblk}
\usepackage{algorithmic}
\usepackage{graphicx}
\usepackage{textcomp}
\usepackage{xcolor}
\usepackage{hyperref}
\usepackage{multirow}
\usepackage{color}
\usepackage{listings}
\usepackage{tablefootnote}
\usepackage{float}
\usepackage[acronym, shortcuts]{glossaries}
\usepackage[numbers]{natbib}
\usepackage{subcaption}
\usepackage{hhline}
\usepackage{array}
\usepackage{todonotes}
\usepackage{makecell}
\newfloat{listing}{tbp}{lol}
\floatname{listing}{Listing}
\def\BibTeX{{\rm B\kern-.05em{\sc i\kern-.025em b}\kern-.08em
    T\kern-.1667em\lower.7ex\hbox{E}\kern-.125emX}}

\makeglossaries

\newacronym[long=artificial intelligence]{AI}{AI}{Artificial Intelligence}
\newacronym[long=machine learning]{ML}{ML}{Machine Learning}
\newacronym[long=Autism spectrum disorder]{ASD}{ASD}{Autism Spectrum Disorder}
\newacronym{ADOS}{ADOS}{Autism Diagnostic Observation Scale}
\newacronym{BLE}{BLE}{Bluetooth Low Energy}
\newacronym[long=domain-specific language]{DSL}{DSL}{Domain-Specific Language}
\newacronym{GATT}{GATT}{Generic Attribute Profile}
\newacronym[long=developmental delay]{DD}{DD}{Developmental Delay}
\newacronym[long=electroencephalogram]{EEG}{EEG}{electroencephalogram}
\newacronym[long=event-related brain potential]{ERP}{ERP}{Event-Related Brain Potential}
\newacronym[plural=HICs,longplural=high-income countries,long=high-income country]{HIC}{HIC}{High-Income Country}
\newacronym[long=intellectual disability]{ID}{ID}{Intellectual Disability}
\newacronym[plural=LMICs,longplural=low- and middle-income countries,long=low- and middle-income country]{LMIC}{LMIC}{Low- and Middle-Income Country}
\newacronym{M-CHAT}{M-CHAT}{Modified Checklist for Autism in Toddlers}
\newacronym{M-CHAT-R/F}{M-CHAT-R/F}{Modified Checklist for Autism in Toddlers, Revised with Follow-Up}
\newacronym{MDAT}{MDAT}{Malawi Developmental Assessment Tool}
\newacronym{PHR}{PHR}{Personal Health Record}
\newacronym{EHR}{EHR}{Electronic Health Record}
\newacronym[long=augmented-reality]{AR}{AR}{Augmented-Reality}
\newacronym[long=software development kit]{SDK}{SDK}{Software Development Kit}
\newacronym{WHO}{WHO}{World Health Organization}
\newacronym{fNIRS}{fNIRS}{Functional Near-Infrared Spectroscopy}
\newacronym[long=rapid eye movement]{REM}{REM}{Rapid Eye Movement}
\newacronym{EDF}{EDF}{European Data Format}
\newacronym{BDF}{BDF}{BioSemi Data Format}
\newacronym[long=analysis object model]{AOM}{AOM}{Analysis Object Model}
\newacronym{PROM}{PROM}{Patient-Reported Outcome Measure}

\glsdisablehyper

\begin{document}

\makeatletter
\newcommand{\linebreakand}{%
  \end{@IEEEauthorhalign}
  \hfill\mbox{}\par
  \mbox{}\hfill\begin{@IEEEauthorhalign}
}
\makeatother

\title{\vspace{-1cm}Toward Scalable Access to Neurodevelopmental Screening: Insights, Implementation, and Challenges}

\author[1]{Andreas Bauer~\orcidlink{0000-0002-1680-237X}~}
\author[2,3]{William Bosl~\orcidlink{0000-0002-8490-0190}~}
\author[1]{Oliver Aalami~\orcidlink{0009-0001-5934-2078}~}
\author[1,*]{Paul Schmiedmayer~\orcidlink{0000-0002-8607-9148}~}

\affil[1]{Stanford Mussallem Center for Biodesign,\protect\\Stanford, CA, 94305, United States}
\affil[2]{Boston Children's Hospital and Harvard Medical School,\protect\\Boston, MA, 02115, United States}
\affil[3]{University of San Francisco, School of Nursing and Health Professions, San Francisco, CA, 94117, United States}
\affil[*]{Corresponding Author: schmiedmayer@stanford.edu}

\maketitle

\begin{abstract}
Children with neurodevelopmental disorders require timely intervention to improve long-term outcomes, yet early screening remains inaccessible in many regions.
A scalable solution integrating standardized assessments with physiological data collection, such as \gls{EEG} recordings, could enable early detection in routine settings by non-specialists.

To address this, we introduce NeuroNest, a mobile and cloud-based platform for large-scale \gls{EEG} data collection, neurodevelopmental screening, and research.  
We provide a comprehensive review of existing behavioral and biomarker-based approaches, consumer-grade \gls{EEG} devices, and emerging machine learning techniques.  
NeuroNest integrates low-cost \gls{EEG} devices with digital screening tools, establishing a scalable, open-source infrastructure for non-invasive data collection, automated analysis, and interoperability across diverse hardware.  

Beyond the system architecture and reference implementation, we highlight key challenges in \gls{EEG} data standardization, device interoperability, and bridging behavioral and physiological assessments. 
Our findings emphasize the need for future research on standardized data exchange, algorithm validation, and ecosystem development to expand screening accessibility.
By providing an extensible, open-source system, NeuroNest advances machine learning-based early detection while fostering collaboration in screening technologies, clinical applications, and public health.
\end{abstract}
\section{Introduction}\label{sec:introduction}

Neurodevelopmental assessments are essential for the early identification of developmental delays and disorders in young children~\cite{marlow2019_review_screenin_tools_autism, johnson_2007_identification_evaluation_asdf_children}.
Timely screening enables early intervention, improving long-term outcomes~\cite{myers_2007_management_of_children_with_asd, filipek_2000_screening_diagnosis_autism, berlin_1998_effectiveness_early_intervention, hwang_2013_dd_routines_based_early_intervention}. 
However, children in resource-limited regions face higher risks and often lack access to diagnostic services~\cite{who_developmental_difficulties_early_childhood}. 
These conditions are particularly prevalent in \glspl{LMIC}, where neurodevelopmental disorders such as \gls{ID}, \gls{DD}, and \gls{ASD} present lifelong challenges that could be mitigated through early detection and intervention~\cite{marlow2019_review_screenin_tools_autism}.

Early intervention, including educational programs and family support, enhances developmental potential~\cite{myers_2007_management_of_children_with_asd}, but its effectiveness depends on timely screening.
Current practice includes autism-specific screening at 18- and 24-month well-baby checkups, as recommended by the American Academy of Pediatrics~\cite{hyman2020executive}.
The \gls{M-CHAT}~\cite{robins2001_modified_checklist_autims_toddlers, robins2014_modified_checklist_autims_toddlers_follow_up} is widely used for \gls{ASD} screening from 18 months onward, with broad adoption in the United States and internationally~\cite{Canal-Bedia2011_mchat_spain, inada_2011_mchat_japan}.
However, these screening practices rely on behavioral assessments, which are time-consuming, require specialists, and are not always reliable for very young children~\cite{marlow2019_review_screenin_tools_autism}. 
Moreover, these tools are largely validated on Western populations and are often unavailable in \glspl{LMIC}, highlighting the need for efficient, objective screening approaches in routine healthcare settings.

Beyond behavioral assessments, exploring biomarkers offers the potential for earlier detection of \gls{ASD} and other neurodevelopmental conditions~\cite{bosl2018_eeg_autism_detection}.
Research suggests that nonlinear \gls{EEG} analysis and supervised tensor factorization~\cite{lock2018_supervised_multiway_factorization} provide a potential cost-effective means of assessing brain function for neurodevelopmental screening~\cite{bosl2018_eeg_autism_detection, sathyanarayana2020_nonlinear_analysis, sathyanarayana2022_measuring_medication_effects, bosl2023_biomarker}.
Scalable, cost-effective surveillance and screening tools are crucial for developmental monitoring, guiding early intervention strategies~\cite{ali2013_developmental_delay_need_tools, zwaigenbaum2015_recommendations_autism_screening}.

Scaling these assessments requires low-cost, widely accessible technology~\cite{gladstone2017_low_resource_neurodevelopment_measuring}. 
Consumer-grade \gls{EEG} devices are emerging as viable tools for clinical applications~\cite{niso_2023_wireless_eeg, sawangjai2020_review_consumer_eeg_sensors, larocco_2020_drowsiness_available_lowcost_egg_headsets}, with major companies exploring \gls{EEG} integration in consumer devices~\cite{apple2023_patent_biosignal_sensing_device}. 
Historically, \gls{EEG} has been underutilized outside epilepsy clinics due to its reliance on visual analysis, but new computational techniques, including AI-driven methods, show promise for early neurodevelopmental disorder detection~\cite{bosl2018_eeg_autism_detection}. 
Current solutions are device-specific~\cite{krigolson2017_choosing_muse, krigolson2021_muse_mobile_brain_performance} or rely on proprietary datasets, limiting scalability.

To develop effective, accessible screening tools, a robust, scalable, and open platform is needed to collect, standardize, and analyze \gls{EEG} data alongside traditional screening methods. 
Large-scale data collection is critical for training machine learning models and clinically validating mobile neurodevelopmental screening systems. 

\begin{figure}
    \centering
    \includegraphics[width=1\linewidth, angle=0]{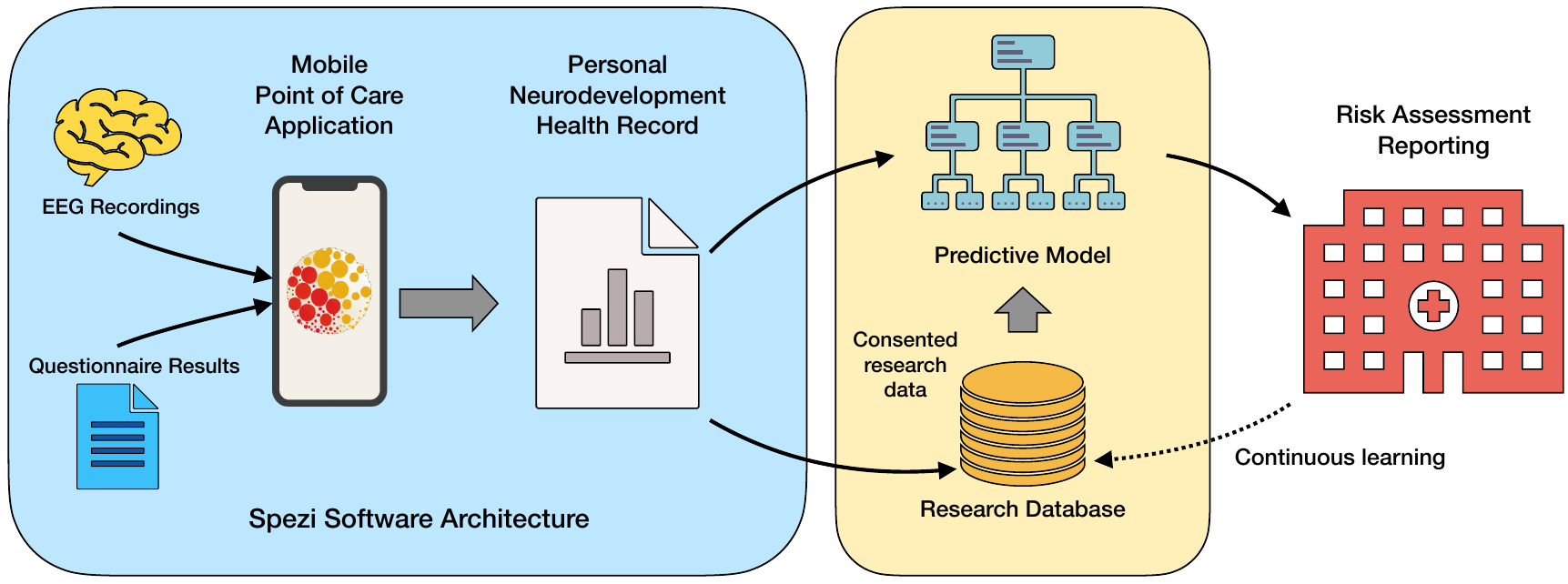}
    \caption[A Platform for Neurodevelopmental Data Collection.]{
        Illustration of a platform for neurodevelopmental data collection.
        The blue-shaded region highlights the key components of this research: a mobile application designed to collect 
        individual patient data, including (1) patient identifiers and demographic information for the \gls{PHR}, 
        (2) \gls{EEG} recordings from a locally connected Bluetooth device, and (3) data from additional sources such as standardized questionnaires.
        The \gls{PHR} seamlessly integrates with algorithmic and AI-based assessment and learning systems.
    }
    \label{fig:project-chart}
\end{figure}

The NeuroNest project (\autoref{fig:project-chart}) addresses these challenges by developing a mobile point-of-care application for neurophysiological and behavioral data collection. 
This system is intended to enable non-specialist community health workers to conduct screenings and \gls{EEG} recordings, with collected data forwarded to a cloud-based component for future automated analysis and probabilistic risk assessment.

This pilot project seeks to demonstrate the feasibility of integrating low-cost, custom-built \gls{EEG} devices into a digital health ecosystem. 
If successful, NeuroNest will provide a scalable platform for early neurodevelopmental screening, supporting non-specialist health workers in both high- and low-resource settings. 
The system will lay the foundation for large-scale clinical testing of screening protocols, initially focusing on intellectual disability, autism, and epilepsy—conditions frequently comorbid with neurodevelopmental disorders~\cite{katchanov2012_epilepsy_care_guidelines, anand2005_epilepsy_clinical_case_definition}. 
By enabling early detection, it will create opportunities for timely intervention.

\section{Related Work}
\label{sec:related-work}

The development of NeuroNest builds on multiple research domains, including neurodevelopmental screening, established assessment tools, emerging technologies, and the feasibility of low-cost \gls{EEG} devices in clinical research.

\subsection{Neurodevelopmental Screening and Early Diagnosis}\label{subsec:neurodevelopment-in-children}

The American Academy of Pediatrics recommends autism screening at 18 and 24 months, alongside general developmental screening starting at nine months~\cite{hyman2020executive}.
These guidelines reflect the current reliance on observable behavioral indicators for \gls{ASD} screening, which are not clinically useful until at least 18 months, when \gls{ASD}-related behaviors begin to emerge~\cite{manner2023prevalence, mccarty2020early, wolff2021predicting}. 
Nevertheless, evidence suggests that earlier intervention leads to better long-term outcomes~\cite{hyman2020identification}.
Thus, a critical challenge in the field is identifying and treating affected children as early as possible~\cite{wolff2021predicting}.

There is growing consensus that atypical neural development precedes behavioral symptoms, beginning shortly after birth or even prenatally~\cite{wen2019first, yin2019brain, liu2024brain}.
Moreover, most available screening tools are either culturally biased toward \glspl{HIC} or prohibitively expensive for widespread use in low-income settings.
Because these tools are developed and validated within specific cultural and socioeconomic contexts, they often fail to generalize effectively to different settings.
As noted by \citeauthor{gladstone2017_low_resource_neurodevelopment_measuring}, \textit{``cross-cultural differences in concepts, norms, beliefs, and values for children’s behaviour are considerable''}~\cite{gladstone2017_low_resource_neurodevelopment_measuring}, especially in \glspl{LMIC}.
These challenges highlight the need for scalable, algorithm-driven approaches to support large-scale clinical testing and the identification of neurodevelopmental biomarkers~\cite{ewen2019conceptual}.
To address this, an infrastructure capable of collecting and managing large-scale data that reflects the cultural and contextual diversity of deployment settings is essential.

\subsection{Behavioral Screening Tools}
\label{subsec:screening-tools}

\citeauthor{marlow2019_review_screenin_tools_autism}~\cite{marlow2019_review_screenin_tools_autism} reviewed 99 screening tools for \gls{ASD} and \gls{DD}, finding that while 35 of 59 \gls{DD} tools were designed for \glspl{LMIC}, only six of 40 \gls{ASD} tools were developed for these regions.
Many tools require licensed professionals or costly proprietary licenses, limiting their scalability in low-resource settings~\cite{marlow2019_review_screenin_tools_autism}.
To address these barriers, \citeauthor{marlow2019_review_screenin_tools_autism} proposed feasibility criteria for \gls{LMIC}-appropriate screening tools: they should take under 30 minutes to administer, cover multiple developmental domains, be freely accessible and low-cost to implement, be usable by community health workers, and demonstrated success or adaptability across multiple \glspl{LMIC}.
The \gls{M-CHAT-R/F} and \gls{MDAT} meet these criteria for \gls{ASD} and \gls{DD}, respectively.
\citeauthor{robins2014_modified_checklist_autims_toddlers_follow_up}~\cite{robins2014_modified_checklist_autims_toddlers_follow_up} demonstrated that the \gls{M-CHAT-R/F}, a two-stage questionnaire, reduces the age of \gls{ASD} diagnosis by up to two years. 
Its classification into low, medium, and high risk enables efficient follow-up, improving early identification rates.

\subsection{Digital and Technology-Enabled Screening}
\label{subsec:technology-enabled-screening}

Digitalizing the screening process enhances accuracy, efficiency, and follow-up compliance. 
\citeauthor{campbell_2017_digital_mchat}~\cite{campbell_2017_digital_mchat} found that a digital \gls{M-CHAT-R/F} improved accurate documentation of results from 54\% to 92\% and improved provider adherence, increasing accurate documentation from 54\% to 92\%.
\citeauthor{campbell_2020_impact_digital_mchat}~\cite{campbell_2020_impact_digital_mchat} showed that digital screening increased the likelihood of referral for evaluation fivefold before 48 months.

Beyond structured questionnaires, mobile-based behavioral assessments have shown promise. 
\citeauthor{egger_2018_researchKit_emotion_attention_analysis_asd}~\cite{egger_2018_researchKit_emotion_attention_analysis_asd} developed a ResearchKit-based app using computer vision to analyze social-emotional behaviors in toddlers at home. 
This approach demonstrated the feasibility of large-scale, automated behavioral screening, reducing reliance on in-person evaluations~\cite{hashemi_carpenter_2015_autism_app, campbell_2018_computer_vision_autism_toddlers}.

\subsection{EEG Biomarkers for Neurodevelopmental Screening}

Behavioral assessment tools have inherent limitations, as atypical brain development likely precedes observable symptoms by months or years~\cite{bosl2018_eeg_autism_detection}. 
\citeauthor{bosl2018_eeg_autism_detection}~\cite{bosl2018_eeg_autism_detection} demonstrated that nonlinear \gls{EEG} analysis can predict \gls{ASD} diagnoses as early as three months of age. 
Their study found that \gls{EEG}-based models correlated with \gls{ADOS} scores at 36 months, suggesting that \gls{EEG} could serve as an early biomarker~\cite{bosl2018_eeg_autism_detection}.
However, the setup by \citeauthor{bosl2018_eeg_autism_detection} relies on a high-density \gls{EEG} setup in controlled conditions, requiring adaptation for real-world use. 
With the increasing availability of low-cost \gls{EEG} devices, there is significant potential to develop accessible, scalable neurodevelopmental screening based on electrophysiological biomarkers~\cite{bosl2018_eeg_autism_detection}.

\subsection{Consumer-Grade EEG Devices}\label{subsec:instantiation-of-consumer-grade-eeg-devices}

Consumer \gls{EEG} devices are becoming more accessible and are increasingly validated for research applications.
\citeauthor{niso_2023_wireless_eeg} reviewed 48 wireless \gls{EEG} devices and 110 studies, highlighting their potential for scalable research~\cite{niso_2023_wireless_eeg}. 
\citeauthor{sawangjai2020_review_consumer_eeg_sensors} compared consumer and research-grade \gls{EEG}, finding that devices from NeuroSky, Emotiv, InteraXon, and OpenBCI showed promise in applications such as cognition, brain-computer interfaces, and education~\cite{sawangjai2020_review_consumer_eeg_sensors}. 

Consumer devices, e.g., the Muse headset has been studied extensively for its feasibility in \gls{EEG}-related research~\cite{bird2018_mental_state_classification_muse, krigolson2017_choosing_muse}.
\citeauthor{krigolson2017_choosing_muse} found that Muse could capture N200, P300, and reward positivity components while reducing setup time from 60 to 10 minutes and lowering costs from \$75,000 to \$250 compared to clinical grade devices~\cite{krigolson2017_choosing_muse}. 
However, they noted challenges such as fixed electrode positions, potential data quality issues, and Bluetooth latency~\cite{krigolson2017_choosing_muse}.

New developments in electrode technology, such as printed, skin-adhesive electrodes, improve usability and data quality. 
\citeauthor{shustak2019_sleep_monitoring} demonstrated that flexible, non-gel electrodes can provide stable wireless \gls{EEG} recordings for sleep monitoring in home settings~\cite{shustak2019_sleep_monitoring}.
\citeauthor{velten2021_portable_electronics_mental_disorders} further developed dry-printed electrodes for \gls{EEG} recording, demonstrating their feasibility in cognitive neuroscience applications~\cite{velten2021_portable_electronics_mental_disorders}. 

\section{Architecture}\label{sec:architecture}

These advancements in low-cost \gls{EEG} technology, digitization, and signal processing pave the way for scalable neurodevelopmental screening.
NeuroNest leverages this progress by integrating consumer \gls{EEG} devices into an open-source platform for large-scale data collection and algorithm development.
\autoref{fig:subsystem-decomposition} illustrates the overall architecture of the NeuroNest platform, which consists of three core subsystems.
At the center of the system is the point-of-care application, which connects with low-cost wearables to collect measurements and other data points, forwarding them to the cloud infrastructure for storage and subsequent analysis.

\begin{figure}
    \centering
    \includegraphics[width=0.8\linewidth, angle=0]{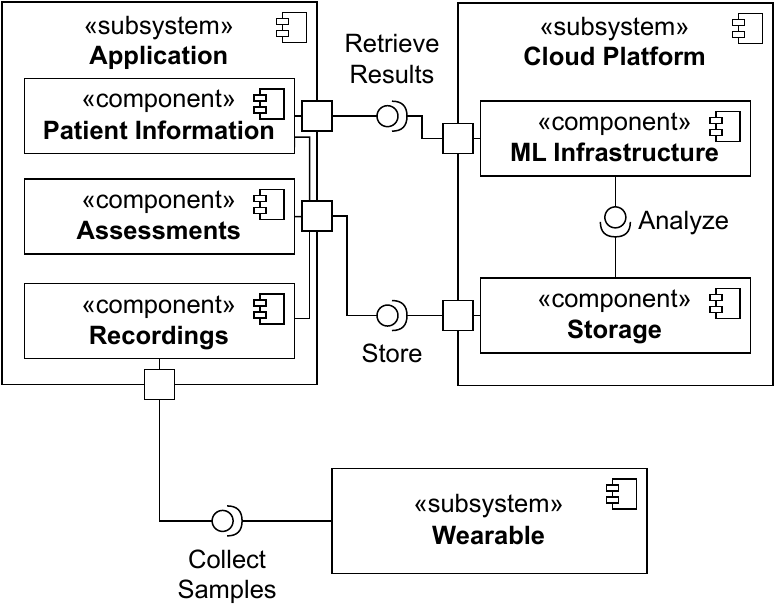}
    \caption{
        Subsystem decomposition of a software architecture for a brain and behavioral data collection 
        platform using wearables, a point-of-care application, and cloud infrastructure.
    }
    \label{fig:subsystem-decomposition}
\end{figure}

\subsection{Wearables}\label{subsec:arch-wearables}

While we primarily consider wearables designed for short-term recordings, the architecture also accommodates devices intended for extended wear. 
The wearable captures high-resolution data during the recording period and transmits it using a local communication protocol. 
For instance, a low-cost \gls{EEG} cap is placed on the patient’s head for the duration of the recording and removed once completed. 
Data samples are streamed in real-time to the application via Bluetooth or a similar wireless connection.

\subsection{Application}\label{subsec:arch-application}

The mobile point-of-care application comprises three primary components: patient management, recording, and assessment functionalities. 
Designed for healthcare providers, the application allows the management of multiple patients while remaining simple enough for use by non-specialist community health workers. 

The patient management module enables users to navigate and oversee a patient database efficiently. 
It includes features to import or add new patient records, manage existing records, and search the patient list. 
Each record stores a unique patient identifier, name, and demographic data, such as date of birth and sex at birth, which can serve as valuable inputs for further analysis.

\autoref{fig:activity-diagram} depicts the high-level workflow of a care provider interacting with the application. 
During an investigation, the provider selects a patient from the list and establishes a connection to a nearby wearable device, if not already paired. 
The wearable sensor is then mounted onto the patient, and recording begins within the application once proper placement is ensured. 
As data is collected, the wearable streams real-time samples to the application, where live visualization enables immediate feedback, helping resolve potential data quality issues. 
The application stores the recorded samples in standardized formats for further analysis.

\begin{figure}
    \centering
    \includegraphics[width=0.75\linewidth, angle=0]{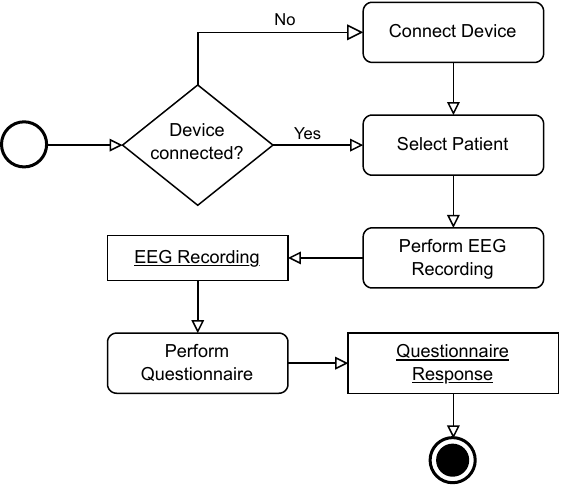}
    \caption{
        Activity diagram describing the high-level interaction of a care provider with the NeuroNest application when 
        collecting data for a given patient.
    }
    \label{fig:activity-diagram}
\end{figure}

Beyond \gls{EEG} data collection, the application supports additional data sources through its assessment module. 
Standardized questionnaires can supplement investigations by providing additional structured data points. 
Digital implementations of these questionnaires streamline assessments, allowing non-specialist personnel to conduct them efficiently while ensuring data consistency and scalability for automated analysis.

\subsection{Cloud Platform}\label{subsec:arch-cloud-platform}

The cloud-based infrastructure facilitates scalable data collection and analysis. 
Upon completion of an investigation, the application securely stores recordings and assessments in standardized formats within the cloud storage system. 
This design supports an extensive data pipeline for analyzing collected data.

Leveraging cloud computing, the platform can dynamically scale computational resources to accommodate machine-learning-based analyses, including computationally intensive and long-running tasks. 
Furthermore, the modular architecture supports manual and repeated evaluations, such as those required in research studies, providing flexibility for various types of analyses and data collection needs.

Additionally, the system enables direct reporting of computational results to either the patient or healthcare provider. 
This functionality extends to risk assessments and diagnostic insights, ensuring timely and relevant information sharing within the healthcare ecosystem.

\section{Results}\label{sec:results}

We designed NeuroNest as a simple, standalone application with the versatility to accommodate various \gls{EEG} devices, ensuring flexibility in its core design. 
It provides a robust framework for conducting research studies aimed at refining and validating neurodevelopmental screening methodologies. 
The platform's effectiveness is measured by its ability to collect data at scale, demonstrating its adaptability and scalability.

NeuroNest~\cite{bauer2023_nams} establishes the foundation for a mobile, cost-effective neurodevelopmental assessment system, enabling large-scale data collection and cloud-supported analysis. 
It is an open-source project licensed under MIT, developed transparently to encourage community collaboration. 
The application integrates the modularized Stanford Spezi software architecture~\cite{schmiedmayer_spezi, schmiedmayer_spzi_template}, emphasizing maintainability and extensibility.

\subsection{Wearable Integration}\label{subsec:results-wearables}

NeuroNest currently supports two mobile \gls{EEG} devices, summarized in \autoref{tab:eeg-devices}.

Muse is a low-cost, consumer-oriented \gls{EEG} headset designed for meditation and sleep tracking~\cite{krigolson2017_choosing_muse, krigolson2021_muse_mobile_brain_performance, bird2018_mental_state_classification_muse}. 
The Muse 2 features a fixed configuration of dry electrodes and collects data from four channels at TP9, AF7, AF9, and TP10, with a maximum sampling frequency of 256 Hz and 12-bit resolution. 
It transmits data via Bluetooth LE using a proprietary, undocumented protocol. 
An \gls{SDK} is available for iOS and Android, but access requires approval\footnote{
    While NeuroNest components are open-source under MIT, the Muse \gls{SDK} is not publicly available. 
    For details, visit \url{https://choosemuse.com/pages/developers}.
}.

SensoMedical’s BIOPOT 3 is a customizable \gls{EEG} and bio-impedance acquisition platform~\cite{velten2021_portable_electronics_mental_disorders, shustak2019_sleep_monitoring}. 
It serves as a signal amplifier supporting an arbitrary 8-channel electrode setup and provides dry and wet electrode caps for flexible placement. 
The device samples at a frequency of 250 Hz, 500 Hz, 1000 Hz, or 2000 Hz with a resolution of 24 bits, transmitting data in real-time via a Bluetooth LE connection.

NeuroNest handles data transmission using the SpeziBluetooth~\cite{schmiedmayer_bluetooth} framework, which extends Apple's CoreBluetooth\footnote{\url{https://developer.apple.com/documentation/corebluetooth}} with an abstraction layer for modern programming paradigms. 
SpeziBluetooth introduces a declarative, composable approach to modeling Bluetooth peripherals, closely mirroring the \gls{GATT} architecture, where services and characteristics define device functionality. 

\begin{table}[h!]
    \centering
    \begin{tabular}{|m{0.22\linewidth}||c|c|}
        \hline
        \textbf{Device} & \textbf{Muse 2} & \textbf{BIOPOT 3} \\
        \hhline{|=#=|=|}
        \textbf{Sensors} & Dry Electrodes & Dry \& Wet Electrodes \\
        \hline
        \textbf{Form Factor} & Rigid headband & Cap \\
        \hline
        \textbf{Size Options} & Fixed & Variable caps \\
        \hline
        \textbf{Channels} & 4 & 8 \\
        \hline
        \textbf{Placement} & TP9, AF7, AF9, TP10 & variable \\
        \hline
        \textbf{Sample\newline Frequencies} & 256 Hz & \makecell{250 Hz, 500 Hz,\\ 1000 Hz, 2000 Hz} \\
        \hline
        \textbf{Resolution} & 12 Bit &  24 Bit \\
        \hline
        \textbf{Protocol} & BLE, Proprietary Service & BLE, Proprietary Service \\
        \hline
        \textbf{File Format} & Proprietary & EDF+/BDF+ \\
        \hline
    \end{tabular}
    \caption{Specifications of supported \gls{EEG} devices.}
    \label{tab:eeg-devices}
\end{table}

\subsection{Application}\label{subsec:results-application}

\newlength{\namsscreenshots}
\setlength{\namsscreenshots}{0.32\linewidth}

\begin{figure}
    \centering
    \begin{subfigure}[t]{0.46\linewidth}
        \centering
        \includegraphics[width=\namsscreenshots]{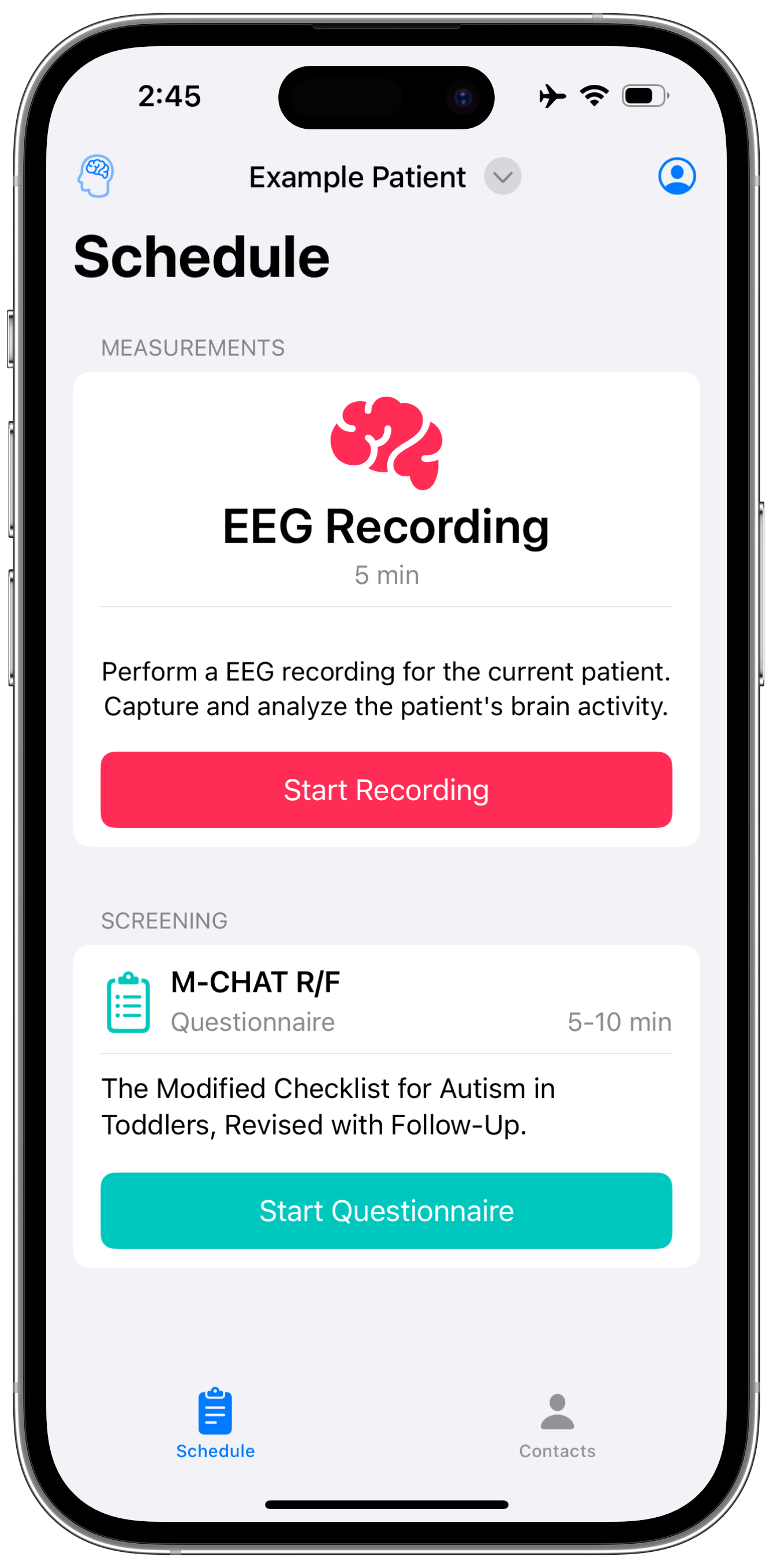}
        \caption{
            Schedule view for the currently selected patient including upcoming measurements and screening tasks.
        }
        \label{fig:neuronest-home}
    \end{subfigure}
    \hfill
    \begin{subfigure}[t]{0.46\linewidth}
        \centering
        \includegraphics[width=\namsscreenshots]{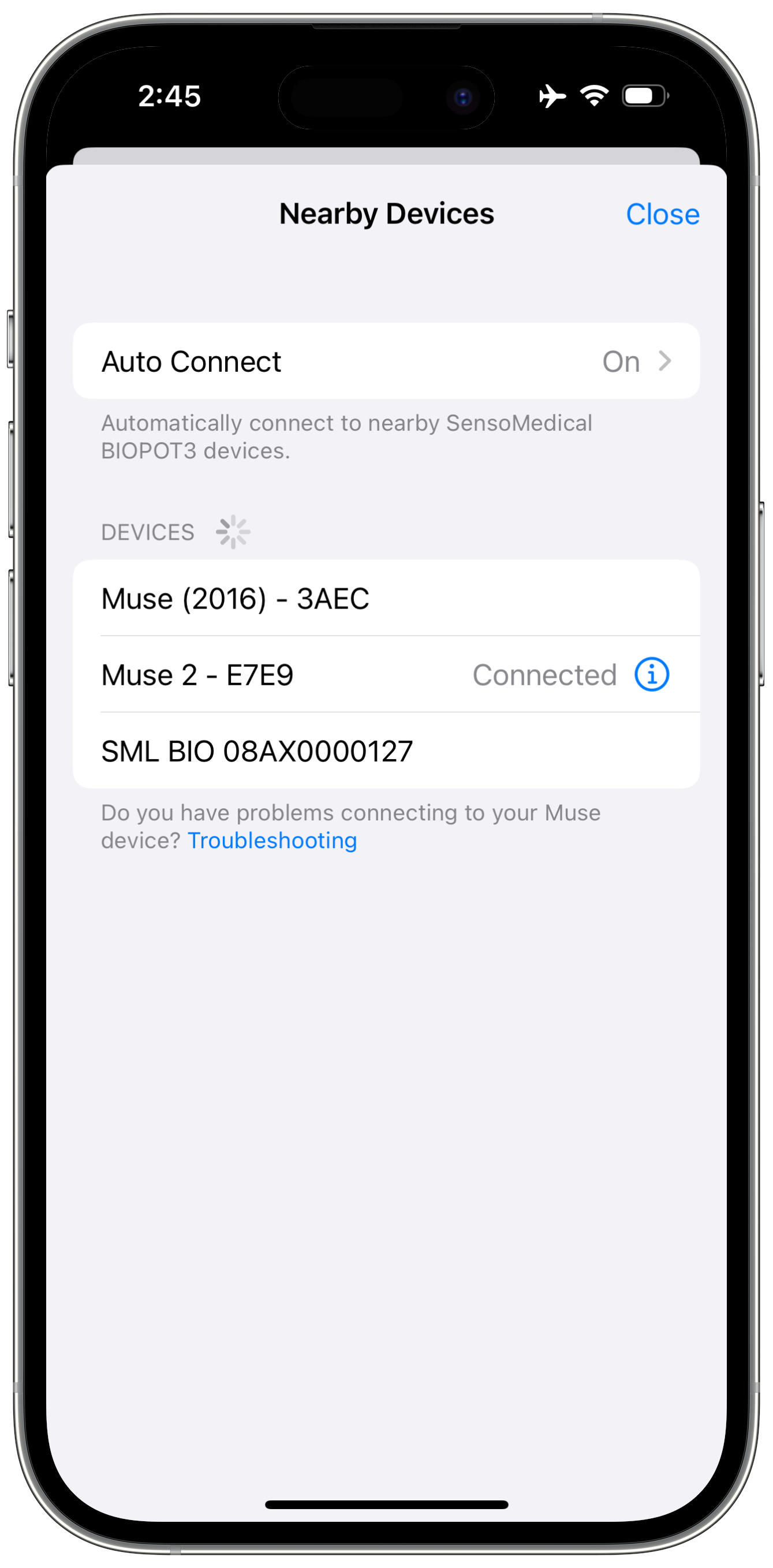}
        \caption{
            Select a device to connect to from the list of nearby devices or view the settings for the currently connected device.
        }
        \label{fig:neuronest-nearbydevices}
    \end{subfigure}
    \hfill
    \begin{subfigure}[t]{0.46\linewidth}
        \centering
        \includegraphics[width=\namsscreenshots]{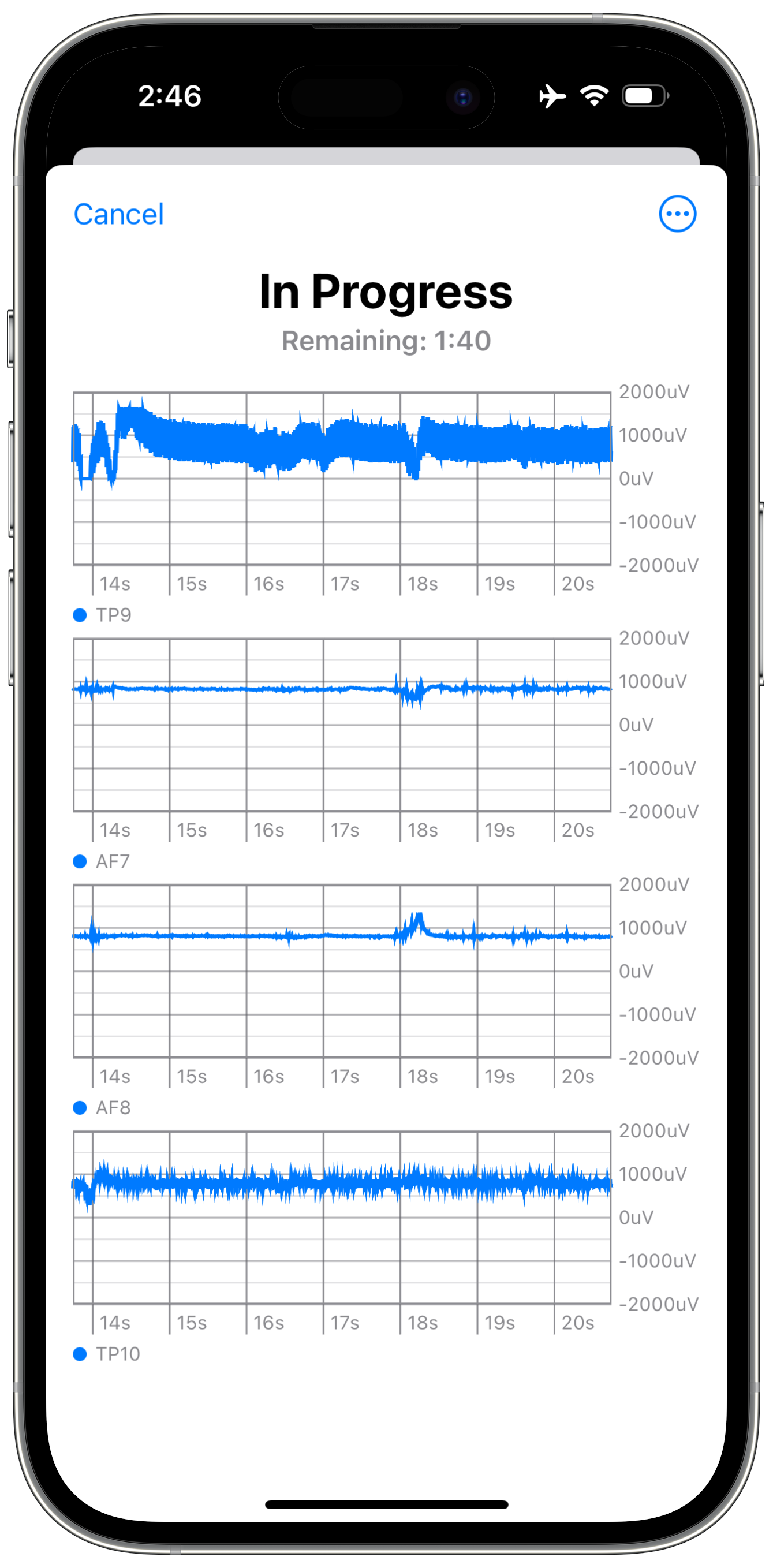}
        \caption{Live visualization of an ongoing \gls{EEG} recording.}
        \label{fig:neuronest-recording}
    \end{subfigure}
    \hfill
    \begin{subfigure}[t]{0.46\linewidth}
        \centering
        \includegraphics[width=\namsscreenshots]{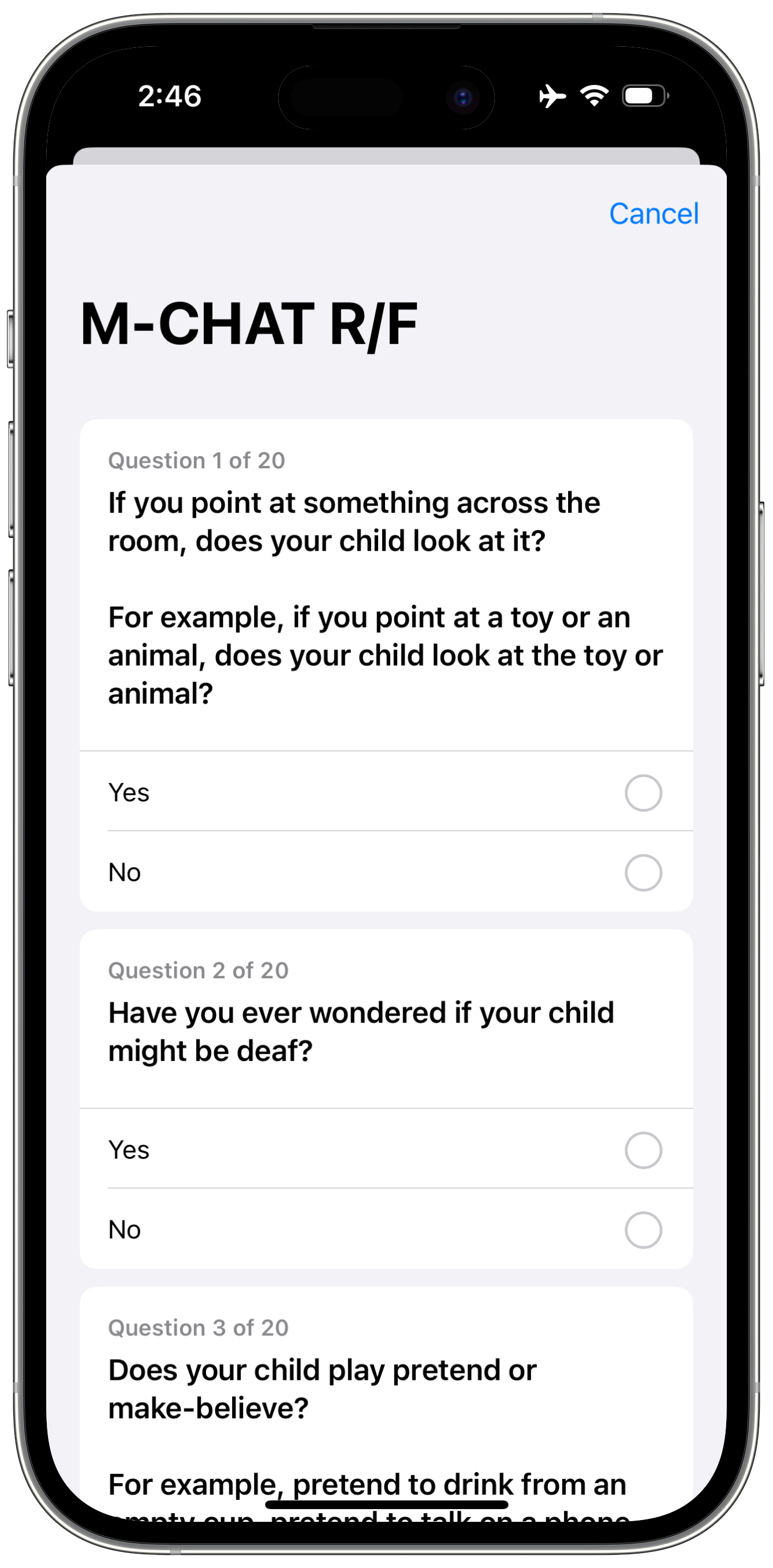}
        \caption{The M-CHAT R/F questionnaire.}
        \label{fig:neuronest-questionnaire}
    \end{subfigure}
    \hfill
    \caption[Screenshots of the NeuroNest iOS Application.]{Screenshots of the NeuroNest iOS application.}
    \label{fig:neuronest-application}
\end{figure}

\autoref{fig:neuronest-application} provides an overview of the NeuroNest iOS application. 
The \textit{Schedule} view (\autoref{fig:neuronest-home}) lists upcoming measurements and screening tasks for the selected patient. 
Patients can be managed via a dedicated interface (\autoref{fig:neuronest-patient-management}), allowing searches, new entries, and selection updates.

To conduct measurements, the application connects to a nearby \gls{EEG} device. 
If not connected automatically, a search can be initiated (\autoref{fig:neuronest-nearbydevices}). 
Device settings can also be adjusted, such as reviewing the Muse headband fit or assigning electrode placements for BIOPOT 3. 
SpeziBluetooth’s modular architecture enables support for multiple Bluetooth peripherals, ensuring extensibility.

During an \gls{EEG} recording session (\autoref{fig:neuronest-recording}), real-time data streaming provides instant feedback to healthcare providers, improving data quality monitoring. 
The application stores waveform data in the standardized \gls{BDF}+~\cite{kemp2003_edf+} format, which extends \gls{EDF} with 24-bit resolution for enhanced precision. 
SpeziFileFormats~\cite{schmiedmayer_file_formats} ensures robust encoding.

To supplement \gls{EEG} data, NeuroNest integrates established screening tools like \gls{M-CHAT-R/F}~(\autoref{fig:neuronest-questionnaire}). 
It leverages \textit{HL7 Structured Data Capture}\footnote{\url{https://hl7.org/fhir/uv/sdc/STU3}} and supports structured questionnaire formats using \textit{FHIR Questionnaire}\footnote{\url{https://hl7.org/fhir/R4/questionnaire.html}} resources. 
Responses are stored in the \textit{FHIR QuestionnaireResponse}\footnote{\url{https://hl7.org/fhir/R4/questionnaireresponse.html}} format, rendered using ResearchKitOnFHIR~\cite{ravi_researchkitonfhir, schmiedmayer_speziquestionnaire}.

\begin{figure}
    \centering
    \begin{subfigure}[t]{0.32\linewidth}
        \centering
        \includegraphics[width=\textwidth]{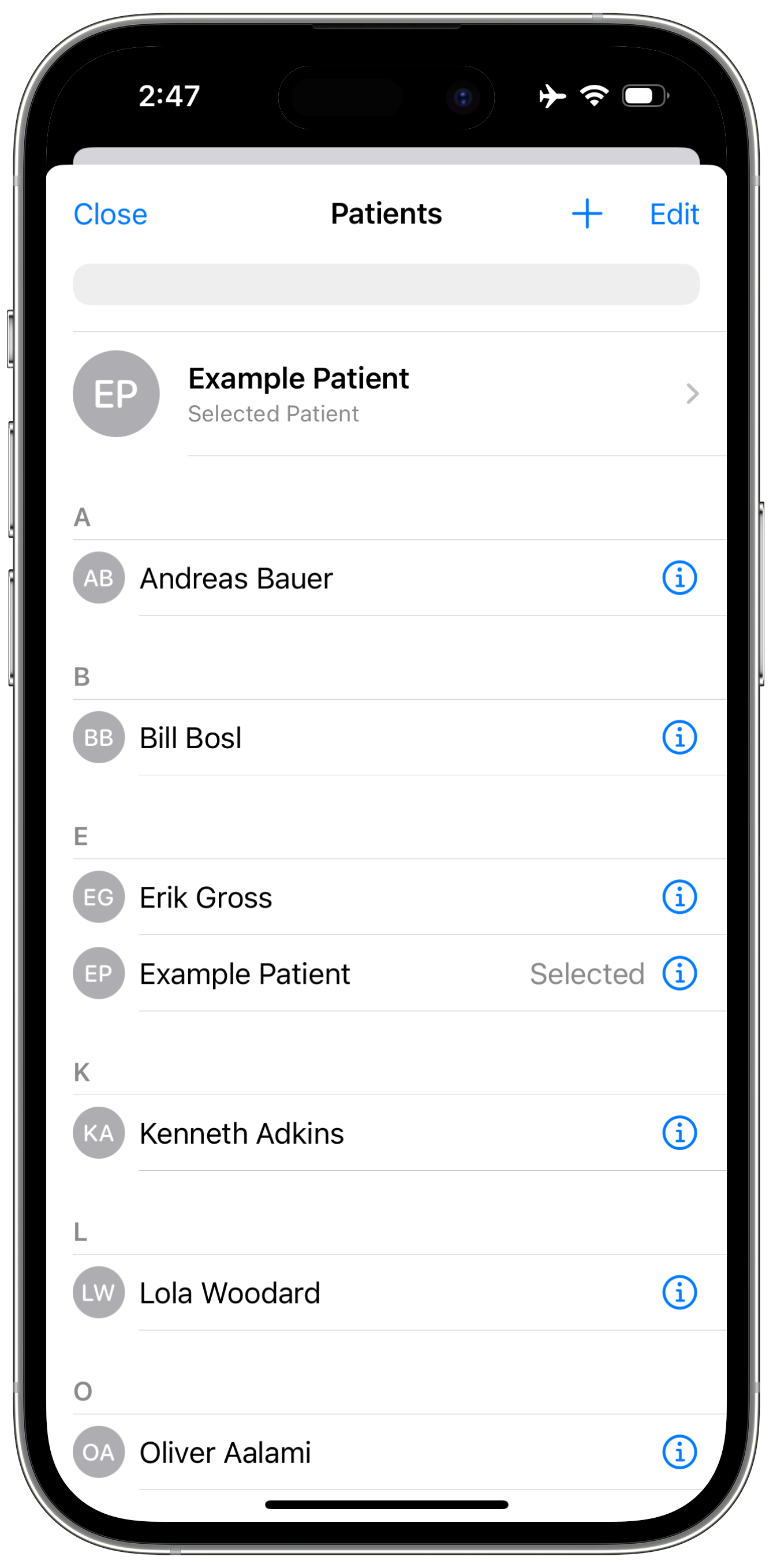}
        \caption{
            List of patients and selected patient.
        }
        \label{fig:neuronest-patientlist}
    \end{subfigure}
    \hfill
    \begin{subfigure}[t]{0.32\linewidth}
        \centering
        \includegraphics[width=\textwidth]{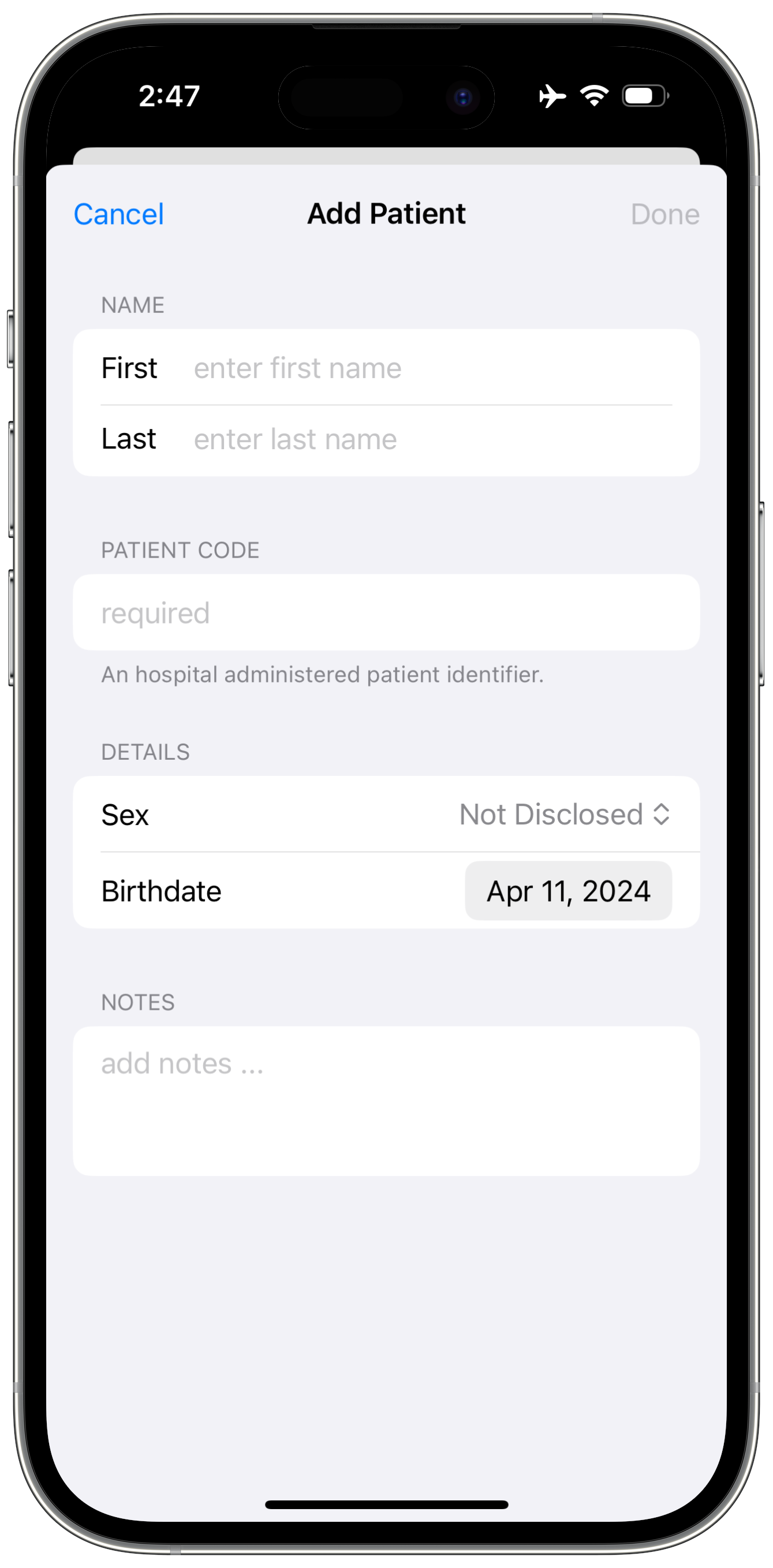}
        \caption{
            Form to add a new patient record.
        }
        \label{fig:neuronest-addpatient}
    \end{subfigure}
    \hfill
    \begin{subfigure}[t]{0.32\linewidth}
        \centering
        \includegraphics[width=\textwidth]{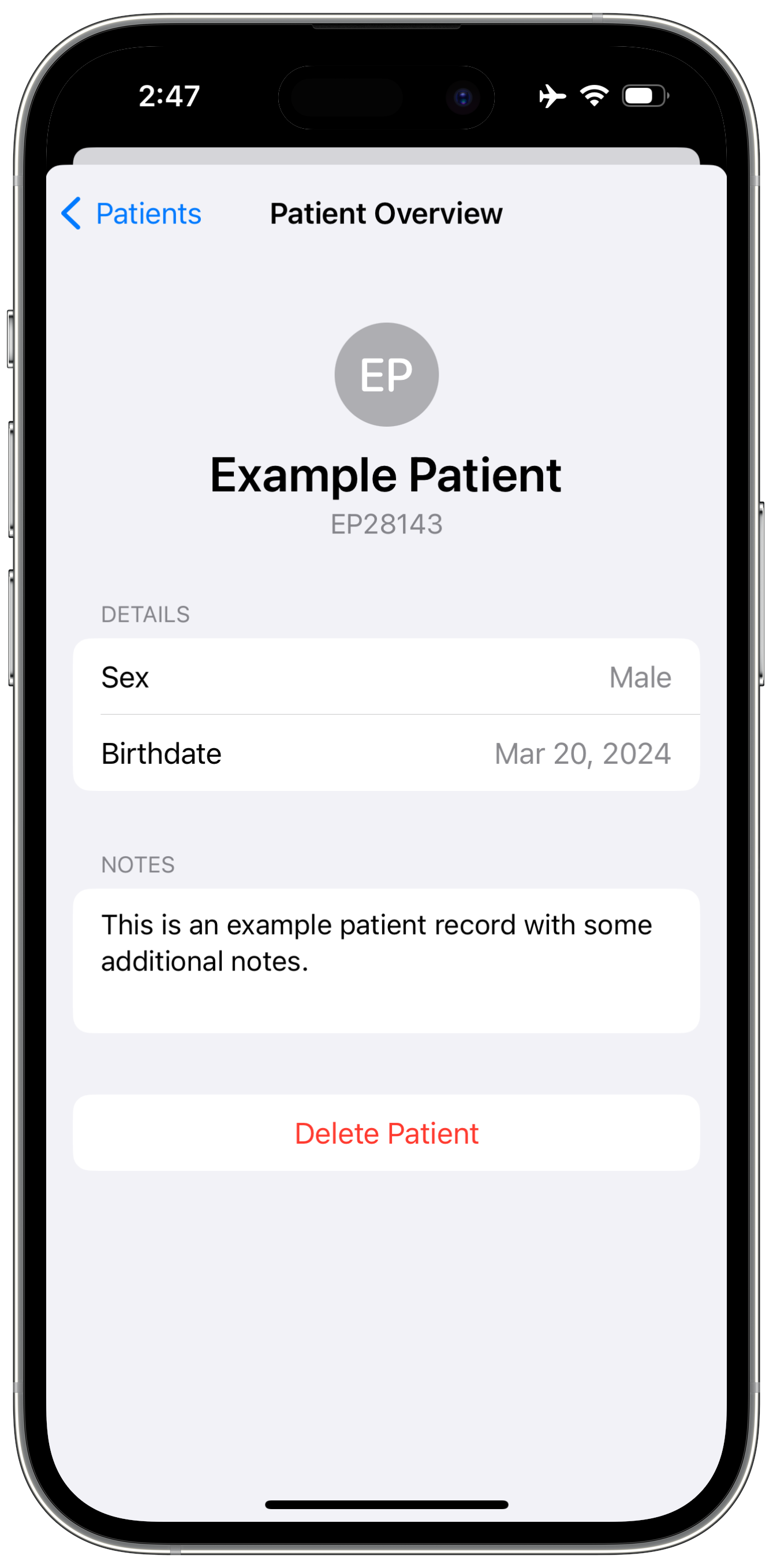}
        \caption{
            Patient Details of the example patient.
        }
        \label{fig:neuronest-examplepatient}
    \end{subfigure}
    \hfill
    \caption[Screenshots of the Patient Management Functionaltities of the NeuroNest iOS Application.]{
        Screenshots of the patient management functionalities of the NeuroNest iOS Application.
    }
    \label{fig:neuronest-patient-management}
\end{figure}

\subsection{Cloud Platform}\label{subsec:results-cloud-platform}

NeuroNest utilizes Google Firebase\footnote{\url{https://firebase.google.com}}, incorporating Authentication, Firestore Database, and Storage. 
Provider authentication ensures secure access to patient data and \gls{EEG} recordings. 
Patient records are stored in Firestore, with questionnaire responses saved in FHIR format for seamless export and integration with external systems.

\gls{EEG} recordings in \gls{BDF}+ format are uploaded to Firebase Storage and linked to corresponding patient records. 
These self-contained files allow easy data sharing with authorized external systems. 
Google Cloud enables dynamic resource allocation for machine-learning-based computations on recorded samples, supporting intensive and long-running analytical tasks.

This modular infrastructure enhances NeuroNest’s capability for large-scale neurodevelopmental data collection, analysis, and research applications.

\section{Discussion}\label{sec:discussion}

The goal of \gls{EEG}-based biomarker research is to identify objective measures of neurodevelopmental trajectories, enabling early detection and intervention for disorders such as epilepsy and autism~\cite{myers_2007_management_of_children_with_asd, johnson_2007_identification_evaluation_asdf_children, berlin_1998_effectiveness_early_intervention}. 
We aimed to evaluate the feasibility and efficacy of a novel platform for \gls{EEG} data collection using an iOS application connected to low-cost \gls{EEG} headsets. 
This platform not only facilitates non-invasive \gls{EEG} recordings in infants but also enables scalable storage and analysis for neurodevelopmental assessments. 
However, implementing \gls{EEG} technology as a screening tool for infants at scale presents unique challenges, including device selection, storage formats, and data analysis methodologies.

\subsection{EEG Device Considerations}

The \gls{EEG} devices integrated with NeuroNest are designed for distinct use cases, each presenting advantages and limitations in research applications. 
The Muse 2 is a consumer-grade headset with a rigid structure, a limited number of channels, and fixed electrode placement~\cite{krigolson2017_choosing_muse, krigolson2021_muse_mobile_brain_performance}.
It poses challenges related to compatibility with different head sizes—particularly for infants—and lacks adaptability for varied recording scenarios~\cite{sawangjai2020_review_consumer_eeg_sensors}.

In contrast, the BIOPOT 3 is a hardware development platform requiring custom configurations rather than being a readily available consumer product~\cite{shustak2019_sleep_monitoring}. 
Its flexibility allows for tailored electrode placements suited to specific research needs. 
Additionally, it offers higher sampling frequencies and resolution (\autoref{tab:eeg-devices}). 
However, its use as a scalable solution requires significant development resources, as it is not available as an off-the-shelf consumer device.

Despite these challenges, consumer-grade \gls{EEG} devices are becoming increasingly accessible and are actively validated in diverse research applications~\cite{sawangjai2020_review_consumer_eeg_sensors, niso_2023_wireless_eeg, larocco_2020_drowsiness_available_lowcost_egg_headsets}. 
Their affordability and ease of use lower the barrier for \gls{EEG} research, expanding potential applications beyond controlled laboratory environments.

\subsection{Challenges in Bluetooth Standardization}

A common issue among all integrated \gls{EEG} devices is the reliance on proprietary Bluetooth communication protocols due to the absence of a standardized \gls{EEG} transmission service. 
While Muse offers an \gls{SDK} for iOS and Android, access is restricted, requiring an application process. 
Furthermore, it does not provide direct specifications for its Bluetooth service. 
In contrast, the BIOPOT 3 Development Kit grants full access to its Bluetooth specification but limits its \gls{SDK} and sample applications to the Android platform. 
This inconsistency introduces additional overhead when integrating \gls{EEG} devices, requiring adaptation to vendor-specific implementations.

Currently, Bluetooth standardization exists for various medical device types, such as blood pressure\footnote{\url{https://www.bluetooth.com/specifications/specs/blood-pressure-service-1-1-1}}, blood glucose\footnote{\url{https://www.bluetooth.com/specifications/specs/glucose-service-1-0-1}}, and weight measurement\footnote{\url{https://www.bluetooth.com/specifications/specs/weight-scale-service-1-0}}. 
However, \gls{EEG} data collection remains outside these efforts, forcing manufacturers to develop proprietary communication protocols. 
This lack of standardization increases integration complexity and limits interoperability across different devices.

While significant efforts have been made to standardize neurophysiology data storage and exchange, some manufacturers still rely on proprietary solutions. 
We strongly encourage collaboration between device manufacturers, the Bluetooth standards organization, and researchers to develop a shared \gls{EEG} Bluetooth specification. 
Such a standard would significantly improve integration, reduce development overhead, and enhance cross-device compatibility.

\subsection{File Format Considerations}

The choice of file storage format is critical for supporting efficient data analysis. 
\gls{EDF}/\gls{EDF}+~\cite{kemp1992_edf, kemp2003_edf+} is a well-established format for neurophysiology signal data, widely supported by open-source tools. 
Its \gls{EDF}+ and \gls{BDF} extensions enhance metadata structuring and increase resolution, respectively.

Despite its widespread adoption, \gls{EDF} lacks certain advanced capabilities, such as synchronized video support, higher data resolution, standardized medical terminology integration, and robust encryption for personal health information~\cite{halford2021_dicom_standard}. 
Recent standardization efforts have introduced support for neurophysiology signal data within the DICOM standard, which is extensively used in hospitals for medical imaging~\cite{halford2021_dicom_standard}. 
The DICOM extension offers up to 32-bit resolution, supports up to 64 waveform channels, and integrates ISO/IEEE 11073 medical terminology while maintaining compatibility with hospital infrastructure~\cite{halford2021_dicom_standard}. 

Additionally, the HL7 FHIR \textit{Guide to Resources}\footnote{\url{https://hl7.org/fhir/R4/resourceguide.html}} includes \gls{EEG} as an example of diagnostic test results that can be represented using \textit{FHIR Observations}. 
However, adoption within clinical practice remains limited. 
We found \textit{FHIR Observations} to be inefficient for \gls{EEG} data storage due to its reliance on string-based encoding, which lacks support for buffered reading and writing.

\subsection{Future Directions}

NeuroNest was designed to be accessible, flexible, and extensible, ensuring its applicability across various research and clinical use cases. 
With the exception of the Muse \gls{SDK}, all components of the application are open-source, fostering collaboration and innovation within the scientific community. 
By leveraging widely used technologies such as Firebase, the system is easily adaptable for different deployment scenarios. 
The decision to use \gls{EDF} as a storage format ensures compatibility with established workflows, although future iterations should consider incorporating more advanced standards such as DICOM.

The application is built to support any Bluetooth-enabled \gls{EEG} device, allowing seamless integration of new hardware with minimal system modifications. 
However, broader efforts within the Bluetooth community are necessary to standardize \gls{EEG} communication protocols, simplifying device interoperability and streamlining development efforts.

Although the system includes a foundational patient management module, future work should focus on enhancing integration with existing \gls{EHR} systems to facilitate seamless data exchange. 
Currently, there is no in-app mechanism for reporting analysis results, such as neurodevelopmental risk assessments, directly to patients or care providers. 
Future iterations should enable automated reporting of diagnostic insights into the patient's health record, expanding the application’s clinical utility.

By addressing these challenges and advancing the NeuroNest platform, we aim to create a scalable, interoperable, and clinically relevant solution for neurodevelopmental screening, supporting early diagnosis and intervention efforts.

\section{Conclusion}\label{sec:conclusion}

A robust data collection platform is essential for developing and validating novel algorithms that enable scalable and early neurodevelopmental screening, facilitating timely intervention. 
In this context, we present NeuroNest, an application designed to streamline data collection processes, integrating \gls{EEG} acquisition and established screening tools into a cloud-based platform. 
This project serves as a catalyst for large-scale data collection, empowering researchers to aggregate and analyze \gls{EEG} signals at scale. 
Its open-source and extensible architecture fosters further research and development, paving the way for more advanced screening tools and interventions.

In its current implementation, the application supports basic patient management, allowing for the storage and handling of demographic data. 
It integrates two low-cost \gls{EEG} devices: Muse 2, a consumer-grade \gls{EEG} headset, and BIOPOT 3, a development kit designed for customized \gls{EEG} hardware solutions. 
Despite the absence of a standardized interface for \gls{EEG} devices, the system remains extensible, enabling integration with a diverse range of devices. 
\gls{EEG} recordings are stored in the \gls{EDF}/\gls{BDF} file format within the cloud infrastructure, ensuring compatibility with established research workflows and analytical tools. 
Additionally, the system incorporates HL7 Structured Data Capture to integrate standardized screening tools as supplementary data sources.

NeuroNest represents a promising foundation for the early screening and monitoring of neurodevelopmental disorders, particularly in underserved regions. 
By providing a mobile, low-friction platform for data collection and analysis, NeuroNest has the potential to accelerate research efforts in neurodevelopmental screening and biomarker discovery. 
Further research and collaboration are essential to refine the platform, validate its efficacy, and enhance its clinical applicability. 

With continued development, NeuroNest can play a critical role in advancing early intervention strategies, ultimately improving long-term outcomes for children and their families.

\section*{Acknowledgment}
We thank all Stanford Spezi contributors for maintaining and supporting the open-source ecosystem.

\bibliographystyle{IEEEtranN}
\bibliography{IEEEabrv,bibliography}

\begin{thebibliography}{54}
\providecommand{\natexlab}[1]{#1}
\providecommand{\url}[1]{#1}
\csname url@samestyle\endcsname
\providecommand{\newblock}{\relax}
\providecommand{\bibinfo}[2]{#2}
\providecommand{\BIBentrySTDinterwordspacing}{\spaceskip=0pt\relax}
\providecommand{\BIBentryALTinterwordstretchfactor}{4}
\providecommand{\BIBentryALTinterwordspacing}{\spaceskip=\fontdimen2\font plus
\BIBentryALTinterwordstretchfactor\fontdimen3\font minus \fontdimen4\font\relax}
\providecommand{\BIBforeignlanguage}[2]{{%
\expandafter\ifx\csname l@#1\endcsname\relax
\typeout{** WARNING: IEEEtranN.bst: No hyphenation pattern has been}%
\typeout{** loaded for the language `#1'. Using the pattern for}%
\typeout{** the default language instead.}%
\else
\language=\csname l@#1\endcsname
\fi
#2}}
\providecommand{\BIBdecl}{\relax}
\BIBdecl

\bibitem[Marlow et~al.(2019)Marlow, Servili, and Tomlinson]{marlow2019_review_screenin_tools_autism}
\BIBentryALTinterwordspacing
M.~Marlow, C.~Servili, and M.~Tomlinson, ``A review of screening tools for the identification of autism spectrum disorders and developmental delay in infants and young children: recommendations for use in low- and middle-income countries,'' \emph{Autism Research}, vol.~12, no.~2, pp. 176--199, 2019. [Online]. Available: \url{https://doi.org/10.1002/aur.2033}
\BIBentrySTDinterwordspacing

\bibitem[Johnson et~al.(2007)Johnson, Myers, , and the Council~on Children With~Disabilities]{johnson_2007_identification_evaluation_asdf_children}
\BIBentryALTinterwordspacing
C.~P. Johnson, S.~M. Myers, , and the Council~on Children With~Disabilities, ``{Identification and Evaluation of Children With Autism Spectrum Disorders},'' \emph{Pediatrics}, vol. 120, no.~5, pp. 1183--1215, 11 2007. [Online]. Available: \url{https://doi.org/10.1542/peds.2007-2361}
\BIBentrySTDinterwordspacing

\bibitem[Myers et~al.(2007)Myers, Johnson, and {the Council on Children With Disabilities}]{myers_2007_management_of_children_with_asd}
\BIBentryALTinterwordspacing
S.~M. Myers, C.~P. Johnson, and {the Council on Children With Disabilities}, ``{Management of Children With Autism Spectrum Disorders},'' \emph{Pediatrics}, vol. 120, no.~5, pp. 1162--1182, 11 2007. [Online]. Available: \url{https://doi.org/10.1542/peds.2007-2362}
\BIBentrySTDinterwordspacing

\bibitem[Filipek et~al.(2000)Filipek, Accardo, Ashwal, Baranek, Cook, Dawson, Gordon, Gravel, Johnson, Kallen, Levy, Minshew, Ozonoff, Prizant, Rapin, Rogers, Stone, Teplin, Tuchman, and Volkmar]{filipek_2000_screening_diagnosis_autism}
\BIBentryALTinterwordspacing
P.~Filipek, P.~Accardo, S.~Ashwal, G.~Baranek, E.~Cook, G.~Dawson, B.~Gordon, J.~Gravel, C.~Johnson, R.~Kallen, S.~Levy, N.~Minshew, S.~Ozonoff, B.~Prizant, I.~Rapin, S.~Rogers, W.~Stone, S.~Teplin, R.~Tuchman, and F.~Volkmar, ``Practice parameter: Screening and diagnosis of autism,'' \emph{Neurology}, vol.~55, no.~4, pp. 468--479, 2000. [Online]. Available: \url{https://doi.org/10.1212/WNL.55.4.468}
\BIBentrySTDinterwordspacing

\bibitem[Berlin et~al.(1998)Berlin, Brooks-Gunn, McCarton, and McCormick]{berlin_1998_effectiveness_early_intervention}
\BIBentryALTinterwordspacing
L.~J. Berlin, J.~Brooks-Gunn, C.~McCarton, and M.~C. McCormick, ``The effectiveness of early intervention: Examining risk factors and pathways to enhanced development,'' \emph{Preventive Medicine}, vol.~27, no.~2, pp. 238--245, 1998. [Online]. Available: \url{https://doi.org/10.1006/pmed.1998.0282}
\BIBentrySTDinterwordspacing

\bibitem[Hwang et~al.(2013)Hwang, Chao, and Liu]{hwang_2013_dd_routines_based_early_intervention}
\BIBentryALTinterwordspacing
A.-W. Hwang, M.-Y. Chao, and S.-W. Liu, ``A randomized controlled trial of routines-based early intervention for children with or at risk for developmental delay,'' \emph{Research in Developmental Disabilities}, vol.~34, no.~10, pp. 3112--3123, 2013. [Online]. Available: \url{https://doi.org/10.1016/j.ridd.2013.06.037}
\BIBentrySTDinterwordspacing

\bibitem[Ertem and {World Health Organization}(2012)]{who_developmental_difficulties_early_childhood}
\BIBentryALTinterwordspacing
I.~O. Ertem and {World Health Organization}, \emph{Developmental difficulties in early childhood: prevention, early identification, assessment and intervention in low- and middle-income countries: a review}.\hskip 1em plus 0.5em minus 0.4em\relax World Health Organization, 2012. [Online]. Available: \url{https://iris.who.int/handle/10665/97942}
\BIBentrySTDinterwordspacing

\bibitem[Hyman et~al.(2020{\natexlab{a}})Hyman, Levy, Myers, COUNCIL ON CHILDREN WITH~DISABILITIES, PEDIATRICS, Kuo, Apkon, Brei, Davidson, Davis, Ellerbeck, Hyman, Noritz, Leppert, Stille, Yin, Weitzman, Bauer, Childers, Levine, Peralta-Carcelen, Smith, Blum, String, Baum, Voigt, and Bridgemohan]{hyman2020executive}
\BIBentryALTinterwordspacing
S.~L. Hyman, S.~E. Levy, S.~M. Myers, S.~O.~D. COUNCIL ON CHILDREN WITH~DISABILITIES, B.~PEDIATRICS, D.~Kuo, S.~Apkon, T.~Brei, L.~F. Davidson, B.~E. Davis, K.~A. Ellerbeck, S.~L. Hyman, G.~H. Noritz, M.~O. Leppert, C.~Stille, L.~Yin, C.~C. Weitzman, N.~S. Bauer, J.~Childers, David~O., J.~M. Levine, A.~M. Peralta-Carcelen, P.~J. Smith, N.~L. Blum, K.~L. String, R.~Baum, R.~Voigt, and C.~Bridgemohan, ``Executive summary: Identification, evaluation, and management of children with autism spectrum disorder,'' \emph{Pediatrics}, vol. 145, no.~1, p. e20193448, 01 2020. [Online]. Available: \url{https://doi.org/10.1542/peds.2019-3448}
\BIBentrySTDinterwordspacing

\bibitem[Robins et~al.(2001)Robins, Fein, Barton, and Green]{robins2001_modified_checklist_autims_toddlers}
\BIBentryALTinterwordspacing
D.~L. Robins, D.~Fein, M.~L. Barton, and J.~A. Green, ``The modified checklist for autism in toddlers: An initial study investigating the early detection of autism and pervasive developmental disorders,'' \emph{Journal of Autism and Developmental Disorders}, vol.~31, no.~2, pp. 131--144, Apr 2001. [Online]. Available: \url{https://doi.org/10.1023/A:1010738829569}
\BIBentrySTDinterwordspacing

\bibitem[Robins et~al.(2014)Robins, Casagrande, Barton, Chen, Dumont-Mathieu, and Fein]{robins2014_modified_checklist_autims_toddlers_follow_up}
\BIBentryALTinterwordspacing
D.~L. Robins, K.~Casagrande, M.~Barton, C.-M.~A. Chen, T.~Dumont-Mathieu, and D.~Fein, ``{Validation of the Modified Checklist for Autism in Toddlers, Revised With Follow-up (M-CHAT-R/F)},'' \emph{Pediatrics}, vol. 133, no.~1, pp. 37--45, 01 2014. [Online]. Available: \url{https://doi.org/10.1542/peds.2013-1813}
\BIBentrySTDinterwordspacing

\bibitem[Canal-Bedia et~al.(2011)Canal-Bedia, Garc{\'\i}a-Primo, Mart{\'\i}n-Cilleros, Santos-Borbujo, Guisuraga-Fern{\'a}ndez, Herr{\'a}ez-Garc{\'\i}a, del Mar Herr{\'a}ez-Garc{\'\i}a, Boada-Mu{\~{n}}oz, Fuentes-Biggi, and Posada-de~la Paz]{Canal-Bedia2011_mchat_spain}
\BIBentryALTinterwordspacing
R.~Canal-Bedia, P.~Garc{\'\i}a-Primo, M.~V. Mart{\'\i}n-Cilleros, J.~Santos-Borbujo, Z.~Guisuraga-Fern{\'a}ndez, L.~Herr{\'a}ez-Garc{\'\i}a, M.~del Mar Herr{\'a}ez-Garc{\'\i}a, L.~Boada-Mu{\~{n}}oz, J.~Fuentes-Biggi, and M.~Posada-de~la Paz, ``Modified checklist for autism in toddlers: Cross-cultural adaptation and validation in spain,'' \emph{Journal of Autism and Developmental Disorders}, vol.~41, no.~10, pp. 1342--1351, Oct 2011. [Online]. Available: \url{https://doi.org/10.1007/s10803-010-1163-z}
\BIBentrySTDinterwordspacing

\bibitem[Inada et~al.(2011)Inada, Koyama, Inokuchi, Kuroda, and Kamio]{inada_2011_mchat_japan}
\BIBentryALTinterwordspacing
N.~Inada, T.~Koyama, E.~Inokuchi, M.~Kuroda, and Y.~Kamio, ``Reliability and validity of the japanese version of the modified checklist for autism in toddlers (m-chat),'' \emph{Research in Autism Spectrum Disorders}, vol.~5, no.~1, pp. 330--336, 2011. [Online]. Available: \url{https://doi.org/10.1016/j.rasd.2010.04.016}
\BIBentrySTDinterwordspacing

\bibitem[Bosl et~al.(2018)Bosl, Tager-Flusberg, and Nelson]{bosl2018_eeg_autism_detection}
\BIBentryALTinterwordspacing
W.~J. Bosl, H.~Tager-Flusberg, and C.~A. Nelson, ``Eeg analytics for early detection of autism spectrum disorder: A data-driven approach,'' \emph{Scientific Reports}, vol.~8, no.~1, p. 6828, May 2018. [Online]. Available: \url{https://doi.org/10.1038/s41598-018-24318-x}
\BIBentrySTDinterwordspacing

\bibitem[Lock and Li(2018)]{lock2018_supervised_multiway_factorization}
\BIBentryALTinterwordspacing
E.~F. Lock and G.~Li, ``{Supervised multiway factorization},'' \emph{Electronic Journal of Statistics}, vol.~12, no.~1, pp. 1150 -- 1180, 2018. [Online]. Available: \url{https://doi.org/10.1214/18-EJS1421}
\BIBentrySTDinterwordspacing

\bibitem[Sathyanarayana et~al.(2020)Sathyanarayana, El~Atrache, Jackson, Alter, Mandl, Loddenkemper, and Bosl]{sathyanarayana2020_nonlinear_analysis}
\BIBentryALTinterwordspacing
A.~Sathyanarayana, R.~El~Atrache, M.~Jackson, A.~S. Alter, K.~D. Mandl, T.~Loddenkemper, and W.~J. Bosl, ``Nonlinear analysis of visually normal eegs to differentiate benign childhood epilepsy with centrotemporal spikes (bects),'' \emph{Scientific Reports}, vol.~10, no.~1, p. 8419, May 2020. [Online]. Available: \url{https://doi.org/10.1038/s41598-020-65112-y}
\BIBentrySTDinterwordspacing

\bibitem[Sathyanarayana et~al.(2022)Sathyanarayana, El~Atrache, Jackson, Cantley, Reece, Ufongene, Loddenkemper, Mandl, and Bosl]{sathyanarayana2022_measuring_medication_effects}
\BIBentryALTinterwordspacing
A.~Sathyanarayana, R.~El~Atrache, M.~Jackson, S.~Cantley, L.~Reece, C.~Ufongene, T.~Loddenkemper, K.~D. Mandl, and W.~J. Bosl, ``Measuring real-time medication effects from electroencephalography,'' \emph{Journal of Clinical Neurophysiology}, May 2022. [Online]. Available: \url{https://doi.org/10.1097/WNP.0000000000000946}
\BIBentrySTDinterwordspacing

\bibitem[Bosl et~al.(2023)Bosl, Bosquet~Enlow, Lock, and Nelson]{bosl2023_biomarker}
\BIBentryALTinterwordspacing
W.~J. Bosl, M.~Bosquet~Enlow, E.~F. Lock, and C.~A. Nelson, ``A biomarker discovery framework for childhood anxiety,'' \emph{Frontiers in Psychiatry}, vol.~14, 2023. [Online]. Available: \url{https://doi.org/10.3389/fpsyt.2023.1158569}
\BIBentrySTDinterwordspacing

\bibitem[Ali et~al.(2013)Ali, Mustafa, Balaji, and Poornima]{ali2013_developmental_delay_need_tools}
\BIBentryALTinterwordspacing
S.~S. Ali, S.~A. Mustafa, P.~A. Balaji, and S.~Poornima, ``Developmental delay: Need of screening tools for primary care providers,'' \emph{Journal of research in medical sciences : the official journal of Isfahan University of Medical Sciences}, vol.~18, no.~11, p. 1013, November 2013. [Online]. Available: \url{https://pmc.ncbi.nlm.nih.gov/articles/PMC3906778/}
\BIBentrySTDinterwordspacing

\bibitem[Zwaigenbaum et~al.(2015)Zwaigenbaum, Bauman, Fein, Pierce, Buie, Davis, Newschaffer, Robins, Wetherby, Choueiri, Kasari, Stone, Yirmiya, Estes, Hansen, McPartland, Natowicz, Carter, Granpeesheh, Mailloux, Smith~Roley, and Wagner]{zwaigenbaum2015_recommendations_autism_screening}
\BIBentryALTinterwordspacing
L.~Zwaigenbaum, M.~L. Bauman, D.~Fein, K.~Pierce, T.~Buie, P.~A. Davis, C.~Newschaffer, D.~L. Robins, A.~Wetherby, R.~Choueiri, C.~Kasari, W.~L. Stone, N.~Yirmiya, A.~Estes, R.~L. Hansen, J.~C. McPartland, M.~R. Natowicz, A.~Carter, D.~Granpeesheh, Z.~Mailloux, S.~Smith~Roley, and S.~Wagner, ``{Early Screening of Autism Spectrum Disorder: Recommendations for Practice and Research},'' \emph{Pediatrics}, vol. 136, no. Supplement\_1, pp. S41--S59, 10 2015. [Online]. Available: \url{https://doi.org/10.1542/peds.2014-3667D}
\BIBentrySTDinterwordspacing

\bibitem[Gladstone et~al.(2017)Gladstone, Abubakar, Idro, Langfitt, and Newton]{gladstone2017_low_resource_neurodevelopment_measuring}
\BIBentryALTinterwordspacing
M.~Gladstone, A.~Abubakar, R.~Idro, J.~Langfitt, and C.~R. Newton, ``Measuring neurodevelopment in low-resource settings,'' \emph{The Lancet Child {\&} Adolescent Health}, vol.~1, no.~4, pp. 258--259, Dec 2017. [Online]. Available: \url{https://doi.org/10.1016/S2352-4642(17)30117-7}
\BIBentrySTDinterwordspacing

\bibitem[Niso et~al.(2023)Niso, Romero, Moreau, Araujo, and Krol]{niso_2023_wireless_eeg}
\BIBentryALTinterwordspacing
G.~Niso, E.~Romero, J.~T. Moreau, A.~Araujo, and L.~R. Krol, ``Wireless eeg: A survey of systems and studies,'' \emph{NeuroImage}, vol. 269, p. 119774, 2023. [Online]. Available: \url{https://doi.org/10.1016/j.neuroimage.2022.119774}
\BIBentrySTDinterwordspacing

\bibitem[Sawangjai et~al.(2020)Sawangjai, Hompoonsup, Leelaarporn, Kongwudhikunakorn, and Wilaiprasitporn]{sawangjai2020_review_consumer_eeg_sensors}
\BIBentryALTinterwordspacing
P.~Sawangjai, S.~Hompoonsup, P.~Leelaarporn, S.~Kongwudhikunakorn, and T.~Wilaiprasitporn, ``Consumer grade eeg measuring sensors as research tools: A review,'' \emph{IEEE Sensors Journal}, vol.~20, no.~8, pp. 3996--4024, 2020. [Online]. Available: \url{https://doi.org/10.1109/JSEN.2019.2962874}
\BIBentrySTDinterwordspacing

\bibitem[LaRocco et~al.(2020)LaRocco, Le, and Paeng]{larocco_2020_drowsiness_available_lowcost_egg_headsets}
\BIBentryALTinterwordspacing
J.~LaRocco, M.~D. Le, and D.-G. Paeng, ``A systemic review of available low-cost eeg headsets used for drowsiness detection,'' \emph{Frontiers in Neuroinformatics}, vol.~14, 2020. [Online]. Available: \url{https://doi.org/10.3389/fninf.2020.553352}
\BIBentrySTDinterwordspacing

\bibitem[Azemi et~al.(2023)Azemi, Moin, Pragada, Lu, Powell, Minxha, and Hotelling]{apple2023_patent_biosignal_sensing_device}
\BIBentryALTinterwordspacing
E.~Azemi, A.~Moin, A.~Pragada, J.~H.-C. Lu, V.~M. Powell, J.~Minxha, and S.~P. Hotelling, ``Biosignal sensing device using dynamic selection of electrodes,'' Patent, July 20, 2023. [Online]. Available: \url{https://patentcenter.uspto.gov/applications/18094841}
\BIBentrySTDinterwordspacing

\bibitem[Krigolson et~al.(2017)Krigolson, Williams, Norton, Hassall, and Colino]{krigolson2017_choosing_muse}
\BIBentryALTinterwordspacing
O.~E. Krigolson, C.~C. Williams, A.~Norton, C.~D. Hassall, and F.~L. Colino, ``Choosing muse: Validation of a low-cost, portable eeg system for erp research,'' \emph{Frontiers in Neuroscience}, vol.~11, 2017. [Online]. Available: \url{https://doi.org/10.3389/fnins.2017.00109}
\BIBentrySTDinterwordspacing

\bibitem[Krigolson et~al.(2021)Krigolson, Hammerstrom, Abimbola, Trska, Wright, Hecker, and Binsted]{krigolson2021_muse_mobile_brain_performance}
\BIBentryALTinterwordspacing
O.~E. Krigolson, M.~R. Hammerstrom, W.~Abimbola, R.~Trska, B.~W. Wright, K.~G. Hecker, and G.~Binsted, ``Using muse: Rapid mobile assessment of brain performance,'' \emph{Frontiers in Neuroscience}, vol.~15, 2021. [Online]. Available: \url{https://doi.org/10.3389/fnins.2021.634147}
\BIBentrySTDinterwordspacing

\bibitem[Katchanov and Birbeck(2012)]{katchanov2012_epilepsy_care_guidelines}
\BIBentryALTinterwordspacing
J.~Katchanov and G.~L. Birbeck, ``Epilepsy care guidelines for low- and middle- income countries: From who mental health gap to national programs,'' \emph{BMC Medicine}, vol.~10, no.~1, p. 107, Sep 2012. [Online]. Available: \url{https://doi.org/10.1186/1741-7015-10-107}
\BIBentrySTDinterwordspacing

\bibitem[Anand et~al.(2005)Anand, Jain, Paul, Srivastava, Sahariah, and Kapoor]{anand2005_epilepsy_clinical_case_definition}
\BIBentryALTinterwordspacing
K.~Anand, S.~Jain, E.~Paul, A.~Srivastava, S.~A. Sahariah, and S.~K. Kapoor, ``Development of a validated clinical case definition of generalized tonic--clonic seizures for use by community-based health care providers,'' \emph{Epilepsia}, vol.~46, no.~5, pp. 743--750, 2005. [Online]. Available: \url{https://doi.org/10.1111/j.1528-1167.2005.41104.x}
\BIBentrySTDinterwordspacing

\bibitem[Maenner et~al.(2023)Maenner, Warren, Williams, Amoakohene, Bakian, Bilder, Durkin, Fitzgerald, Furnier, Hughes, Ladd-Acosta, McArthur, Pas, Salinas, Vehorn, Williams, Esler, Grzybowski, Hall-Lande, Nguyen, Pierce, Zahorodny, Hudson, Hallas, Mancilla, Patrick, Shenouda, Sidwell, DiRienzo, Gutierrez, Spivey, Lopez, Pettygrove, Schwenk, Washington, and Shaw]{manner2023prevalence}
\BIBentryALTinterwordspacing
M.~J. Maenner, Z.~Warren, A.~R. Williams, E.~Amoakohene, A.~V. Bakian, D.~A. Bilder, M.~S. Durkin, R.~T. Fitzgerald, S.~M. Furnier, M.~M. Hughes, C.~M. Ladd-Acosta, D.~McArthur, E.~T. Pas, A.~Salinas, A.~Vehorn, S.~Williams, A.~Esler, A.~Grzybowski, J.~Hall-Lande, R.~H.~N. Nguyen, K.~Pierce, W.~Zahorodny, A.~Hudson, L.~Hallas, K.~C. Mancilla, M.~Patrick, J.~Shenouda, K.~Sidwell, M.~DiRienzo, J.~Gutierrez, M.~H. Spivey, M.~Lopez, S.~Pettygrove, Y.~D. Schwenk, A.~Washington, and K.~A. Shaw, ``\BIBforeignlanguage{eng}{Prevalence and characteristics of autism spectrum disorder among children aged 8 years - autism and developmental disabilities monitoring network, 11 sites, united states, 2020.}'' \emph{\BIBforeignlanguage{eng}{MMWR Surveill Summ}}, vol.~72, no.~2, pp. 1--14, Mar 2023. [Online]. Available: \url{https://doi.org/10.15585/mmwr.ss7202a1}
\BIBentrySTDinterwordspacing

\bibitem[McCarty and Frye(2020)]{mccarty2020early}
\BIBentryALTinterwordspacing
P.~McCarty and R.~E. Frye, ``\BIBforeignlanguage{eng}{Early detection and diagnosis of autism spectrum disorder: Why is it so difficult?}'' \emph{\BIBforeignlanguage{eng}{Semin Pediatr Neurol}}, vol.~35, p. 100831, Oct 2020. [Online]. Available: \url{https://doi.org/10.1016/j.spen.2020.100831}
\BIBentrySTDinterwordspacing

\bibitem[Wolff and Piven(2021)]{wolff2021predicting}
\BIBentryALTinterwordspacing
J.~J. Wolff and J.~Piven, ``\BIBforeignlanguage{eng}{Predicting autism in infancy.}'' \emph{\BIBforeignlanguage{eng}{J Am Acad Child Adolesc Psychiatry}}, vol.~60, no.~8, pp. 958--967, Aug 2021. [Online]. Available: \url{https://doi.org/10.1016/j.jaac.2020.07.910}
\BIBentrySTDinterwordspacing

\bibitem[Hyman et~al.(2020{\natexlab{b}})Hyman, Levy, and Myers]{hyman2020identification}
\BIBentryALTinterwordspacing
S.~L. Hyman, S.~E. Levy, and S.~M. Myers, ``\BIBforeignlanguage{eng}{Identification, evaluation, and management of children with autism spectrum disorder.}'' \emph{\BIBforeignlanguage{eng}{Pediatrics}}, vol. 145, no.~1, Jan 2020. [Online]. Available: \url{https://doi.org/10.1542/peds.2019-3447}
\BIBentrySTDinterwordspacing

\bibitem[Wen et~al.(2019)Wen, Zhang, Li, Liu, Yin, Lin, Zhang, and Shen]{wen2019first}
\BIBentryALTinterwordspacing
X.~Wen, H.~Zhang, G.~Li, M.~Liu, W.~Yin, W.~Lin, J.~Zhang, and D.~Shen, ``First-year development of modules and hubs in infant brain functional networks,'' \emph{NeuroImage}, vol. 185, pp. 222--235, 2019. [Online]. Available: \url{https://doi.org/10.1016/j.neuroimage.2018.10.019}
\BIBentrySTDinterwordspacing

\bibitem[Yin et~al.(2019)Yin, Chen, Hung, Baluyot, Li, and Lin]{yin2019brain}
\BIBentryALTinterwordspacing
W.~Yin, M.-H. Chen, S.-C. Hung, K.~R. Baluyot, T.~Li, and W.~Lin, ``Brain functional development separates into three distinct time periods in the first two years of life,'' \emph{NeuroImage}, vol. 189, pp. 715--726, 2019. [Online]. Available: \url{https://doi.org/10.1016/j.neuroimage.2019.01.025}
\BIBentrySTDinterwordspacing

\bibitem[Liu et~al.(2024)Liu, Lu, Kim, Lee, Duffy, Yuan, Chai, Cole, Wu, Toga, Jahanshad, Gano, Barkovich, Xu, and Kim]{liu2024brain}
\BIBentryALTinterwordspacing
M.~Liu, M.~Lu, S.~Y. Kim, H.~J. Lee, B.~A. Duffy, S.~Yuan, Y.~Chai, J.~H. Cole, X.~Wu, A.~W. Toga, N.~Jahanshad, D.~Gano, A.~J. Barkovich, D.~Xu, and H.~Kim, ``Brain age predicted using graph convolutional neural network explains neurodevelopmental trajectory in preterm neonates,'' \emph{European Radiology}, vol.~34, no.~6, pp. 3601--3611, 2024. [Online]. Available: \url{https://doi.org/10.1007/s00330-023-10414-8}
\BIBentrySTDinterwordspacing

\bibitem[Ewen et~al.(2019)Ewen, Sweeney, and Potter]{ewen2019conceptual}
\BIBentryALTinterwordspacing
J.~B. Ewen, J.~A. Sweeney, and W.~Z. Potter, ``Conceptual, regulatory and strategic imperatives in the early days of eeg-based biomarker validation for neurodevelopmental disabilities,'' \emph{Frontiers in integrative neuroscience}, vol.~13, p.~45, 2019. [Online]. Available: \url{https://doi.org/10.3389/fnint.2019.00045}
\BIBentrySTDinterwordspacing

\bibitem[Campbell et~al.(2017)Campbell, Carpenter, Espinosa, Hashemi, Qiu, Tepper, Calderbank, Sapiro, Egger, Baker, and Dawson]{campbell_2017_digital_mchat}
\BIBentryALTinterwordspacing
K.~Campbell, K.~L.~H. Carpenter, S.~Espinosa, J.~Hashemi, Q.~Qiu, M.~Tepper, R.~Calderbank, G.~Sapiro, H.~L. Egger, J.~P. Baker, and G.~Dawson, ``\BIBforeignlanguage{en}{Use of a digital modified checklist for autism in toddlers - revised with follow-up to improve quality of screening for autism},'' \emph{\BIBforeignlanguage{en}{J Pediatr}}, vol. 183, pp. 133--139.e1, Feb. 2017. [Online]. Available: \url{https://doi.org/10.1016/j.jpeds.2017.01.021}
\BIBentrySTDinterwordspacing

\bibitem[Major et~al.(2020)Major, Campbell, Espinosa, Baker, Carpenter, Sapiro, Vermeer, and Dawson]{campbell_2020_impact_digital_mchat}
\BIBentryALTinterwordspacing
S.~Major, K.~Campbell, S.~Espinosa, J.~P. Baker, K.~L. Carpenter, G.~Sapiro, S.~Vermeer, and G.~Dawson, ``\BIBforeignlanguage{en}{Impact of a digital modified checklist for autism in {Toddlers-Revised} on likelihood and age of autism diagnosis and referral for developmental evaluation},'' \emph{\BIBforeignlanguage{en}{Autism}}, vol.~24, no.~7, pp. 1629--1638, May 2020. [Online]. Available: \url{https://doi.org/10.1177/1362361320916656}
\BIBentrySTDinterwordspacing

\bibitem[Egger et~al.(2018)Egger, Dawson, Hashemi, Carpenter, Espinosa, Campbell, Brotkin, Schaich-Borg, Qiu, Tepper, Baker, Bloomfield, and Sapiro]{egger_2018_researchKit_emotion_attention_analysis_asd}
\BIBentryALTinterwordspacing
H.~L. Egger, G.~Dawson, J.~Hashemi, K.~L.~H. Carpenter, S.~Espinosa, K.~Campbell, S.~Brotkin, J.~Schaich-Borg, Q.~Qiu, M.~Tepper, J.~P. Baker, R.~A. Bloomfield, Jr, and G.~Sapiro, ``\BIBforeignlanguage{en}{Automatic emotion and attention analysis of young children at home: a {ResearchKit} autism feasibility study},'' \emph{\BIBforeignlanguage{en}{NPJ Digit Med}}, vol.~1, p.~20, Jun. 2018. [Online]. Available: \url{https://doi.org/10.1038/s41746-018-0024-6}
\BIBentrySTDinterwordspacing

\bibitem[Hashemi et~al.(2015)Hashemi, Campbell, Carpenter, Harris, Qiu, Tepper, Espinosa, Schaich~Borg, Marsan, Calderbank, Baker, Egger, Dawson, and Sapiro]{hashemi_carpenter_2015_autism_app}
\BIBentryALTinterwordspacing
J.~Hashemi, K.~Campbell, K.~Carpenter, A.~Harris, Q.~Qiu, M.~Tepper, S.~Espinosa, J.~Schaich~Borg, S.~Marsan, R.~Calderbank, J.~Baker, H.~Egger, G.~Dawson, and G.~Sapiro, ``A scalable app for measuring autism risk behaviors in young children: A technical validity and feasibility study,'' vol.~3, 01 2015. [Online]. Available: \url{https://doi.org/10.4108/eai.14-10-2015.2261939}
\BIBentrySTDinterwordspacing

\bibitem[Campbell et~al.(2018)Campbell, Carpenter, Hashemi, Espinosa, Marsan, Borg, Chang, Qiu, Vermeer, Adler, Tepper, Egger, Baker, Sapiro, and Dawson]{campbell_2018_computer_vision_autism_toddlers}
\BIBentryALTinterwordspacing
K.~Campbell, K.~L. Carpenter, J.~Hashemi, S.~Espinosa, S.~Marsan, J.~S. Borg, Z.~Chang, Q.~Qiu, S.~Vermeer, E.~Adler, M.~Tepper, H.~L. Egger, J.~P. Baker, G.~Sapiro, and G.~Dawson, ``\BIBforeignlanguage{en}{Computer vision analysis captures atypical attention in toddlers with autism},'' \emph{\BIBforeignlanguage{en}{Autism}}, vol.~23, no.~3, pp. 619--628, Mar. 2018. [Online]. Available: \url{https://doi.org/10.1177/1362361318766247}
\BIBentrySTDinterwordspacing

\bibitem[Bird et~al.(2018)Bird, Manso, Ribeiro, Ek{\'a}rt, and Faria]{bird2018_mental_state_classification_muse}
\BIBentryALTinterwordspacing
J.~J. Bird, L.~J. Manso, E.~P. Ribeiro, A.~Ek{\'a}rt, and D.~R. Faria, ``A study on mental state classification using eeg-based brain-machine interface,'' in \emph{2018 International Conference on Intelligent Systems (IS)}, Sep. 2018, pp. 795--800. [Online]. Available: \url{https://doi.org/10.1109/IS.2018.8710576}
\BIBentrySTDinterwordspacing

\bibitem[Shustak et~al.(2019)Shustak, Inzelberg, Steinberg, Rand, David~Pur, Hillel, Katzav, Fahoum, De~Vos, Mirelman, and Hanein]{shustak2019_sleep_monitoring}
\BIBentryALTinterwordspacing
S.~Shustak, L.~Inzelberg, S.~Steinberg, D.~Rand, M.~David~Pur, I.~Hillel, S.~Katzav, F.~Fahoum, M.~De~Vos, A.~Mirelman, and Y.~Hanein, ``Home monitoring of sleep with a temporary-tattoo eeg, eog and emg electrode array: a feasibility study,'' \emph{Journal of neural engineering}, vol.~16, no.~2, p. 026024, 2019. [Online]. Available: \url{https://doi.org/10.1088/1741-2552/aafa05}
\BIBentrySTDinterwordspacing

\bibitem[Velten et~al.(2021)Velten, Schuck, Knoll, Wagner, Volk, Hanein, Hendler, Farah, and Asfour]{velten2021_portable_electronics_mental_disorders}
\BIBentryALTinterwordspacing
T.~Velten, H.~Schuck, T.~Knoll, S.~Wagner, D.~Volk, Y.~Hanein, T.~Hendler, M.~Farah, and L.~Asfour, \emph{Nano-based portable electronics for the diagnosis of mental disorders and functional restoration, production technologies and devices}, 2021. [Online]. Available: \url{https://doi.org/10.24406/publica-fhg-301341}
\BIBentrySTDinterwordspacing

\bibitem[Bauer et~al.(2025)Bauer, Schmiedmayer, and Aalami]{bauer2023_nams}
\BIBentryALTinterwordspacing
A.~Bauer, P.~Schmiedmayer, and O.~Aalami, ``{Neurodevelopment Assessment and Monitoring System (NAMS)},'' Feb. 2025. [Online]. Available: \url{https://doi.org/10.5281/zenodo.8374397}
\BIBentrySTDinterwordspacing

\bibitem[Schmiedmayer et~al.(2024{\natexlab{a}})Schmiedmayer, Ravi, Aalami, and Bauer]{schmiedmayer_spezi}
\BIBentryALTinterwordspacing
P.~Schmiedmayer, V.~Ravi, O.~Aalami, and A.~Bauer, ``Spezi,'' Mar. 2024. [Online]. Available: \url{https://doi.org/10.5281/zenodo.7538238}
\BIBentrySTDinterwordspacing

\bibitem[Schmiedmayer et~al.(2024{\natexlab{b}})Schmiedmayer, Ravi, and Aalami]{schmiedmayer_spzi_template}
\BIBentryALTinterwordspacing
P.~Schmiedmayer, V.~Ravi, and O.~Aalami, ``Spezi template application,'' Mar. 2024. [Online]. Available: \url{https://doi.org/10.5281/zenodo.7600783}
\BIBentrySTDinterwordspacing

\bibitem[Schmiedmayer and Bauer(2024{\natexlab{a}})]{schmiedmayer_bluetooth}
\BIBentryALTinterwordspacing
P.~Schmiedmayer and A.~Bauer, ``Spezibluetooth,'' Mar. 2024. [Online]. Available: \url{https://doi.org/10.5281/zenodo.10020080}
\BIBentrySTDinterwordspacing

\bibitem[Kemp and Olivan(2003)]{kemp2003_edf+}
\BIBentryALTinterwordspacing
B.~Kemp and J.~Olivan, ``European data format `plus' (edf+), an edf alike standard format for the exchange of physiological data,'' \emph{Clinical Neurophysiology}, vol. 114, no.~9, pp. 1755--1761, 2003. [Online]. Available: \url{https://doi.org/10.1016/S1388-2457(03)00123-8}
\BIBentrySTDinterwordspacing

\bibitem[Schmiedmayer and Bauer(2024{\natexlab{b}})]{schmiedmayer_file_formats}
\BIBentryALTinterwordspacing
P.~Schmiedmayer and A.~Bauer, ``Spezifileformats,'' Mar. 2024. [Online]. Available: \url{https://doi.org/10.5281/zenodo.10724947}
\BIBentrySTDinterwordspacing

\bibitem[Ravi et~al.(2024)Ravi, Schmiedmayer, Aalami, and Bauer]{ravi_researchkitonfhir}
\BIBentryALTinterwordspacing
V.~Ravi, P.~Schmiedmayer, O.~Aalami, and A.~Bauer, ``Researchkitonfhir,'' Mar. 2024. [Online]. Available: \url{https://doi.org/10.5281/zenodo.7538169}
\BIBentrySTDinterwordspacing

\bibitem[Schmiedmayer et~al.(2024{\natexlab{c}})Schmiedmayer, Ravi, and Aalami]{schmiedmayer_speziquestionnaire}
\BIBentryALTinterwordspacing
P.~Schmiedmayer, V.~Ravi, and O.~Aalami, ``Speziquestionnaire,'' Mar. 2024. [Online]. Available: \url{https://doi.org/10.5281/zenodo.7706903}
\BIBentrySTDinterwordspacing

\bibitem[Kemp et~al.(1992)Kemp, V{\"a}rri, Rosa, Nielsen, and Gade]{kemp1992_edf}
\BIBentryALTinterwordspacing
B.~Kemp, A.~V{\"a}rri, A.~C. Rosa, K.~D. Nielsen, and J.~Gade, ``A simple format for exchange of digitized polygraphic recordings,'' \emph{Electroencephalography and Clinical Neurophysiology}, vol.~82, no.~5, pp. 391--393, 1992. [Online]. Available: \url{https://doi.org/10.1016/0013-4694(92)90009-7}
\BIBentrySTDinterwordspacing

\bibitem[Halford et~al.(2021)Halford, Clunie, Brinkmann, Krefting, R{\'e}mi, Rosenow, Husain, F{\"u}rbass, {Andrew Ehrenberg}, and Winkler]{halford2021_dicom_standard}
\BIBentryALTinterwordspacing
J.~J. Halford, D.~A. Clunie, B.~H. Brinkmann, D.~Krefting, J.~R{\'e}mi, F.~Rosenow, A.~Husain, F.~F{\"u}rbass, J.~{Andrew Ehrenberg}, and S.~Winkler, ``Standardization of neurophysiology signal data into the dicom{\textregistered} standard,'' \emph{Clinical Neurophysiology}, vol. 132, no.~4, pp. 993--997, 2021. [Online]. Available: \url{https://doi.org/10.1016/j.clinph.2021.01.019}
\BIBentrySTDinterwordspacing

\end{thebibliography}

\end{document}